\documentclass[prd,aps,eqsecnum,floatfix,nofootinbib,preprint,tightenlines]{revtex4}

\usepackage{latexsym}
\usepackage{graphicx}
\usepackage{multirow}
\usepackage{float}
\usepackage{comment}
\usepackage[dvipsnames]{xcolor}

\def\bibi{\bibitem}

\let\inodot=\i

\def\d{\delta}
\def\f{\phi}                    
\def\g{\gamma}

\def\i{\iota}

\def\l{\lambda}
\def\m{\mu}
\def\n{\nu}

\def\p{\pi}                     
\def\r{\rho}                    
\def\s{\sigma}                  
\def\t{\tau}

\def\D{\Delta}

\def\P{\Pi}

\def\co{{\cal O}}

\def\cbo{{\,\raise-.15ex\Sc [\,}}                       
\def\ltap{\raisebox{-.4ex}{\rlap{$\sim$}} \raisebox{.4ex}{$<$}}   

\def\svev#1{\left\langle #1\right\rangle}       

\def\ddt#1{{\buildrel {\hbox{\LARGE .\kern-2pt.}} \over {#1}}}

\def\ie{\mbox{\it i.e.}}
\def\eg{\mbox{\it e.g.}}
\def\etc{\mbox{\it etc.}}

\def\half{{1\over 2}}

\def\floatcaption#1#2{ \caption{#2 \label{#1}} }

\def\bibi{\bibitem}    

\def\ttl#1{{\it #1}}

\def\seef{{\it cf.\  }}

\begin{document}

\begin{boldmath}
\begin{center}
{\large{\bf
Spectral-weight sum rules for the\\ hadronic vacuum polarization}
}\\[8mm]
Diogo Boito,$^a$ Maarten Golterman,$^{b,c}$ Kim Maltman$^{d,e}$ and
Santiago Peris,$^c$\\[8 mm]
$^a$Instituto de F{\'\inodot}sica de S{\~a}o Carlos, Universidade de S{\~a}o
Paulo\\ CP 369, 13570-970, S{\~a}o Carlos, SP, Brazil
\\[5mm]
$^b$Department of Physics and Astronomy, San Francisco State University,\\
San Francisco, CA 94132, USA
\\[5mm]
$^c$Department of Physics and IFAE-BIST, Universitat Aut\`onoma de Barcelona\\
E-08193 Bellaterra, Barcelona, Spain
\\
[5mm]
$^d$Department of Mathematics and Statistics,
York University\\  Toronto, ON Canada M3J~1P3
\\[5mm]
$^e$CSSM, University of Adelaide, Adelaide, SA~5005 Australia
\\[10mm]
\end{center}
\end{boldmath}
\begin{quotation}
We develop a number of sum rules comparing spectral integrals involving
judiciously chosen weights to integrals over the corresponding Euclidean
two-point function. The applications we have in mind are to the hadronic
vacuum polarization that determines the most important hadronic correction
$a_\m^{\rm HVP}$ to the muon anomalous magnetic moment. First, we point
out how spectral weights may be chosen that emphasize narrow regions in
$\sqrt{s}$, providing a tool to investigate emerging discrepancies between
data-driven and lattice determinations of $a_\m^{\rm HVP}$. Alternatively,
for a narrow region around the $\r$ mass, they may allow for a comparison
of the dispersive determination of $a_\m^{\rm HVP}$ with lattice
determinations zooming in on the region of the well-known BaBar--KLOE
discrepancy. Second, we show how such sum rules can in principle
be used for carrying out precision comparisons of hadronic-$\t$-decay-based
data and $e^+e^-\to\mbox{hadrons}(\g)$-based data, where lattice
computations can provide the necessary isospin-breaking corrections.
\end{quotation}

\newpage
\section{\label{intro} Introduction}
As is well known, the recent FNAL E989 experimental result for the muon
anomalous magnetic moment $a_\m=(g-2)/2$ \cite{FNL} confirms the earlier
BNL E821 result \cite{BNL} and produces an updated experimental world average
$4.2\s$ larger than the $g-2$ Theory Initiative assessment \cite{review} of
the Standard Model (SM) prediction, based on the work of
Ref.~\cite{Aoyama:2012wk,Aoyama:2019ryr,Czarnecki:2002nt,Gnendiger:2013pva,Davier:2017zfy,Keshavarzi:2018mgv,Davier:2019can,KNT19,Colangelo:2018mtw,Hoferichter:2019mqg,Hoid:2020xjs,Kurz:2014wya,Melnikov:2003xd,Masjuan:2017tvw,Colangelo:2017qdm,Colangelo:2017fiz,Hoferichter:2018dmo,Hoferichter:2018kwz,Gerardin:2019vio,Bijnens:2019ghy,Colangelo:2019lpu,Colangelo:2019uex,Colangelo:2014qya,Blum:2019ugy}.
However, a lattice result for the hadronic vacuum polarization (HVP)
contribution $a_\m^{\rm HVP}$ \cite{BMW} comes out $2.1\s$ higher than the
dispersive $R$-ratio-based estimate which underlies the SM value
of Ref.~\cite{review}. Replacing the dispersive estimate for $a_\m^{\rm HVP}$
with the value found in Ref.~\cite{BMW}, the SM-based estimate for $a_\m$ is
only $1.5\s$ lower than the world-average experimental value. As there is
no evidence for discrepancies in other contributions to $a_\m$, the recent
focus has been on understanding the discrepancy between the dispersive and
lattice values for $a_\m^{\rm HVP}$.\footnote{For instance, there is
good agreement between data-driven and lattice values for the hadronic
light-by-light contribution, within $\,\ltap\, 20\%$ errors. This $20\%$
error corresponds to an uncertainty of about $2\times 10^{-10}$ in $a_\m$,
which is not sufficient to explain the discrepancy.}

The dispersive {\it vs.} lattice discrepancy becomes even more pronounced
if we consider the ``intermediate window'' quantity introduced by
RBC/UKQCD \cite{RBC}, in which the integral over Euclidean time, $t$,
of the lattice correlator that yields $a_\m^{\rm HVP}$ is restricted to a
``window'' between $t=0.4$ and $t=1$~fm (smeared by a width of $0.15$~fm
on both boundaries to avoid lattice artifacts) by multiplying the integrand
with a (smoothed-out) double step function in $t$. While the original
RBC/UKQCD study found agreement between the lattice-based result and the
corresponding electro-production-based dispersive estimate,
Ref.~\cite{ABGP19} found a significantly larger lattice value. This larger
value was subsequently confirmed in Ref.~\cite{BMW}, which produced a lattice
result $3.7\s$ above the dispersive estimate. The virtue of the RBC/UKQD
intermediate window quantity is that it can be computed with smaller
errors on the lattice than the quantity $a_\m^{\rm HVP}$ itself, thus
allowing more stringent tests between different lattice computations, as
well as between lattice and dispersive results.

Meanwhile, the larger lattice values for the intermediate window quantity
found in Refs.~\cite{ABGP19,BMW} have been confirmed, with different lattice
discretizations of the QCD action and the electromagnetic (EM) current,
by a number of other groups as well as in updates of the results of
Refs.~\cite{RBC,ABGP19}, in Refs.~\cite{LM20,chiQCD,ABGP22,Mainz,ETMC,RBC22,FHM22}.

Clearly, then, it is important to develop further tools to study the
discrepancies between dispersive and lattice estimates for $a_\m^{\rm HVP}$
and closely related quantities such as the intermediate window. To obtain
dispersive estimates for the window quantity, which is defined as a
function of Euclidean time, the window function needs to be converted
to a window in $\sqrt{s}$, the center-of-mass energy in $e^+e^-\to$~hadrons.
As a function of $\sqrt{s}$, however, the weight which defines the
intermediate window is very broad, ranging from about 0.7~GeV to 3~GeV
(taking the values of $\sqrt{s}$ where it exceeds roughly half of its
 maximum value). To explore the lattice {\it vs.} dispersive discrepancy
in more detail, it would be useful to have access to tools to compare
data-driven dispersive results with lattice results in narrower,
more ``custom-designed'' windows in $\sqrt{s}$. In this paper, we propose
a class of sum rules designed with precisely this goal in mind. Similar
ideas were recently explored in Ref.~\cite{Rwindow} by considering linear
combinations of a set of Euclidean windows, and in Ref.~\cite{TdeG}. Here,
instead, we define the weights we will employ in our weighted spectral
integrals directly as a function of $s$, and derive sum rules relating
these weighted spectral integrals to integrals in Euclidean time over
correlation functions which can be evaluated on the lattice.

In Sec.~\ref{sumrules} we develop two sets of sum rules starting from weights
defined as a function of $s$. In Sec.~\ref{sumrule1} we consider a class of
rational weights that allow us to define windows localized in $s$.
They are similar to those used recently
in Ref.~\cite{Kim} to obtain a lattice-based determination of
$\vert V_{us}\vert$ from strange hadronic $\t$-decay data.
In Sec.~\ref{sumrule2} we use these rational weights as the starting point
for defining a set of sum-of-exponential weights with shapes very
similar to those defined by the underlying rational weights, following
ideas proposed in Ref.~\cite{HLT}. In both cases exact sum rules exist
relating spectral integrals employing these weights to quantities that
can be directly computed on the lattice. We explain why the
sum-of-exponential weights may lead to smaller errors for the
lattice side of the sum rules than the corresponding rational weights,
and provide examples of this reduction in Sec.~\ref{comparison}.

Narrower windows in $\sqrt{s}$ are also potentially useful for
comparing $I=1$ contributions to $a_\m^{\rm HVP}$ inferred from $I=1$
hadronic $\t$-decay data with the corresponding contributions obtained
using $R$-ratio data. An example of the potential usefulness of narrower
windows is the application to the BaBar--KLOE discrepancy in the
two-pion spectral distributions (for a review, see Ref.~\cite{review}),
where the discrepancy occurs over a fairly narrow range in energy
around the $\r$ peak. Attempts to use $\t$-based data have a long
history \cite{Davier1,Davier2,Jegerlehner,Zhang}, but have been abandoned
more recently because of the increased precision of electroproduction data,
and the lack of a solid theoretical framework for evaluating the
isospin-breaking (IB) corrections that must be applied to the $\t$-based
data. It would be interesting to revisit this possibility since (i) a more
precise $\t$-based non-strange vector spectral function is now
available \cite{AO}, and (ii) Belle II may provide improved
$\t$-decay-distribution data for at least some of the most
important vector-channel exclusive modes \cite{BELLE,BELLE2}.
Moreover, we will argue in Sec.~\ref{tau} of this paper that if the
combined $2\p$ and $4\p$ channels are taken from $\t$, then, to
good accuracy, the lattice can be used to compute the necessary
IB corrections from first principles. We illustrate the comparison
of electroproduction- and hadronic-$\t$-decay-based $2\p+4\p$ data
using two rational weight choices, $W_{1,5}$ and $W_{2,5}$,
defined in Sec.~\ref{comparison}.

We end the paper with a brief conclusion in Sec.~\ref{conclusions}, and
relegate some technical details to two appendices.

\section{\label{sumrules} Sum rules}
In this section, we develop two types of sum rules, one, in
Sec.~\ref{sumrule1}, based on a set of weights used before in Ref.~\cite{Kim},
and one, in Sec.~\ref{sumrule2}, based on the ideas advocated in Ref.~\cite{HLT}.
They are closely related, and we will explore their differences in examples
in Sec.~\ref{comparison}.

We start from a Euclidean current-current correlator
\begin{equation}
\label{currcorr}
G_{\m\n}(x)=\svev{j_\m(0) j^\prime_\n(x)}
=\int\frac{d^4q}{(2\p)^4}\,e^{iqx}\left(\d_{\m\n}q^2-q_\m q_\n\right)
\P(q^2)\ ,
\end{equation}
for two potentially different vector currents $j_\m$ and $j^\prime_\n$,
and define from this the time-momentum correlator $C(t)$ by
\begin{equation}
\label{Ct}
C(t)=-\frac{1}{3}\sum_{k=1}^3\int d^3x\svev{j_k(0) j^\prime_k(\vec{x},t)}
=-\int\frac{dQ}{2\p}\,e^{iQt}\,Q^2\P(Q^2)\ ,
\end{equation}
where $Q=q_4$, and we have assumed that a regulator (such as the lattice)
has been introduced, so that $C(t)$ and $\P(Q^2)$ are finite. While in
straightforward applications to $a_\m^{\rm HVP}$ the currents $j_\m$ and
$j_\nu^\prime$ will both be the hadronic electromagnetic current, they can
also be chosen different, as will be done in the application described
in Sec.~\ref{tau} below.  The corresponding subtracted polarization
\begin{equation}
\label{subtract}
\hat\P(Q^2)=\P(Q^2)-\P(0)
\end{equation}
can be expressed in terms of $C(t)$ by
\begin{eqnarray}
\label{FT}
\hat\P(Q^2)&=&-\frac{1}{Q^2}\int_{-\infty}^\infty dt
\left(e^{iQt}-1\right)C(t)-\P(0)\\
&=&-\frac{2}{Q^2}\int_0^\infty dt\left(\cos(Qt)-1\right)C(t)-\P(0)\nonumber\\
&=&\int_0^\infty dt\,\left(\frac{4\sin^2(Qt/2)}{Q^2}-t^2\right)\,C(t)\ ,
\nonumber
\end{eqnarray}
where we used $\int dt\,C(t)=0$ and, in the second step, that $C(t)$ is an
even function of $t$. We define the spectral function $\r(s)$ as usual by
\begin{equation}
\label{sf}
\r(s)=\frac{1}{\p}\,\mbox{Im}\,\P(s)\ ,
\end{equation}
with $\hat\P(Q^2)$ and $\r(s)$ satisfying the subtracted dispersion relation
\begin{equation}
\label{disp}
\hat\P(Q^2)=-Q^2\int_{s_{\rm th}}^\infty ds\,\frac{\r(s)}{s(s+Q^2)}\ ,
\end{equation}
with $s_{\rm th}$ the relevant threshold value for $\r(s)$.

\subsection{\label{sumrule1} Rational-weight sum rules}
\begin{figure}
\vspace*{4ex}
\begin{center}
\includegraphics*[width=6cm]{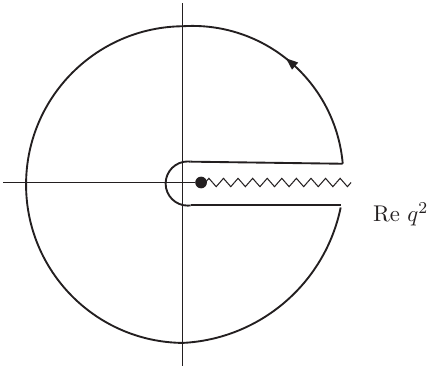}
\end{center}
\begin{quotation}
\floatcaption{cauchy-fig}%
{{\it Contour $C$ used in Eq.~(\ref{cauchy});
$z=q^2=-Q^2$. The black dot indicates the point $q^2=s_{\rm th}$.}}
\label{fig:cauchy}
\end{quotation}
\vspace*{-4ex}
\end{figure}

We begin with a set of spectral weights of the form
\begin{equation}
\label{srweight}
W_{m,n}(s;\{Q_\ell^2\})=\mu^{2(n-m-1)}
\frac{(s-s_{\rm th})^m}{\prod_{\ell=1}^n(s+Q_\ell^2)}\ ,
\qquad Q_n^2>Q_{n-1}^2>\cdots >Q_1^2>0\ .
\end{equation}
Here the $Q_\ell^2$ are a set of fixed Euclidean squared momenta, and we
will always take $m$ sufficiently smaller than $n$ that the weighted
spectral integral with weight~(\ref{srweight}) is finite. We multiply by
a generic mass scale $\mu^{2(n-m-1)}$ to make the weighted spectral
integrals considered below dimensionless; any precisely known scale is
suitable for this purpose and we will employ the choice $\mu=m_\t$ in
what follows.

We then consider the integral
\begin{equation}
\label{cauchy}
\frac{1}{2\p i}\int_C dz\, W_{m,n}(z;\{Q_\ell^2\}) \P(-z)=
(-1)^m m_\t^{2(n-m-1)}\sum_{k=1}^n\frac{(Q_k^2+s_{\rm th})^m}
{\prod_{\ell\ne k}(Q_\ell^2-Q_k^2)}\,\P(Q_k^2)\ ,
\end{equation}
with $C$ the contour in the complex $z=q^2=-Q^2$ plane shown in
Fig.~\ref{cauchy-fig}, assumed to have a radius large enough that
all the $z=-Q_k^2$ points lie in its interior on the negative $z$-axis. The
result on the right-hand side follows from the fact that $\P(-z)$ is
analytic in the complex plane except for a cut starting at $s_{\rm th}$
on the positive real $z$-axis, as shown in Fig. \ref{fig:cauchy}. If we
now take the radius of the circular part of $C$ to infinity, we obtain the
sum rule
\begin{equation}
\label{sumrulemom}
I_{m,n}\equiv\int_{s_{\rm th}}^\infty ds\, W_{m,n}(s;\{Q_\ell^2\})
\r(s)=(-1)^m m_\t^{2(n-m-1)}\sum_{k=1}^n\frac{(Q_k^2+s_{\rm th})^m}
{\prod_{\ell\ne k}(Q_\ell^2-Q_k^2)}\,\P(Q_k^2)\ ,
\end{equation}
where the integral over the spectral function comes from the discontinuity
along the positive real $z$-axis. We will prove in App.~\ref{precision} that
\begin{equation}
\label{sumzero}
\sum_{k=1}^n\frac{(Q_k^2+s_{\rm th})^m}
{\prod_{\ell\ne k}(Q_\ell^2-Q_k^2)} = 0\ ,
\end{equation}
from which it follows that we can replace $\P(Q_k^2)$ by $\hat\P(Q_k^2)$
on the right-hand side of Eq.~(\ref{sumrulemom}). This has to be the case,
as the left-hand side is finite by construction, and thus the term
proportional to $\P(0)$ on the right-hand side has to vanish.

A potentially useful application of this sum rule is to evaluate the
spectral integral using data, for instance obtained from $R$-ratio
measurements, and the sum on the right-hand side using data from lattice
QCD. In fact, what is usually computed in lattice QCD is a position-space
correlator; in the case of $a_\m^{\rm HVP}$, Eq.~(\ref{Ct}) with
$j_\mu =j_\mu^\prime$ the hadronic electromagnetic current.
Replacing $\P(Q_k^2)$ with $\hat\P(Q_k^2)$ and using Eq.~(\ref{FT}), the sum
rule~(\ref{sumrulemom}) can then be recast as
\begin{equation}
\label{sumruleCt}
\int_{s_{\rm th}}^\infty ds\, W_{m,n}(s;\{Q_\ell^2\}) \r(s)=
\int_0^\infty dt\,c^{(m,n)}(t)\,C(t)\ ,
\end{equation}
with
\begin{equation}
\label{cmndef}
c^{(m,n)}(t)=(-1)^m m_\t^{2(n-m-1)}\,\sum_{k=1}^n\,
\frac{(Q_k^2+s_{\rm th})^m}{\prod_{\ell\ne k}(Q_\ell^2-Q_k^2)}
\left(\frac{4\sin^2(Q_kt/2)}{Q_k^2}-t^2\right)\ .
\end{equation}
We will refer to the sum rules of Eq.~~(\ref{sumruleCt}) as
``rational-weight sum rules'' (RWSRs). On the lattice, the integral
over $t$ on the right-hand side of Eq.~(\ref{sumruleCt}) would be replaced
by a sum over the discrete values of $t$ available on the lattice,
using, for example, the trapezoidal rule.
We will discuss examples in Sec.~\ref{exsumrule1} below.

\subsection{\label{sumrule2} Exponential-weight sum rules}

The sum rules of Eq.~(\ref{sumruleCt}) require the evaluation of $C(t)$ at
all values of $t$. In lattice QCD, the signal-to-noise ratio for $C(t)$
deteriorates when $t$ gets large, and the contributions for these
larger $t$ values will thus ``degrade'' the precision with which the
right-hand side of Eq.~(\ref{sumruleCt}) can be computed.

In this section, we therefore develop modified sum rules, involving
new weights constructed to require the evaluation of $C(t)$ at only a
limited set of $t$ values. While these new weights will, of
necessity, differ from the old weights $W_{m,n}$, we will see that,
following the ideas of Ref.~\cite{HLT}, it is possible to choose them
to be remarkably similar to the $W_{m,n}$.

We begin with the observation that
\begin{equation}
\label{Crho}
C(t)=\int_{E_{\rm th}}^\infty dE\,\r(E^2)(E^2\,e^{-E|t|}-2E\,\d(t))\ ,
\end{equation}
with $E_{\rm th}=\sqrt{s_{\rm th}}$ the threshold energy. The term
proportional to $\d(t)$ ensures that $\int dt\, C(t)=0$.\footnote{This
term is formally divergent, but, as before, we assume that $C(t)$ has
been regulated.} We then define new weights
\begin{equation}
\label{wform}
w_n(E;\{t_j\},\{x_j\})=\sum_{j=1}^n x_j(E^2 e^{-|t_j| E}-2E\d(t_j))\ ,
\end{equation}
with $\{ t_j\}$ a fixed set of $t$ values and $\{ x_j\}$ a set of
coefficients to be determined below. In all our applications, the $t_j$
will be chosen positive, and we can ignore the term with the $\d$-function.
The $w_n(E;\{t_j\},\{x_j\})$-weighted spectral integrals then
satisfy sum rules,
\begin{equation}
\label{sumruleexponpre}
\int_{E_{\rm th}}^\infty dE\,\r(E^2)w_n(E;\{t_j\},\{x_j\})
=\sum_{j=1}^n x_j C(t_j)\ ,
\end{equation}
in which only the $n$ values $C(t_j)$ occur on the right-hand side.

The goal then is to start with an initial weight, $W(s)$, having
some desired $s$ dependence (for example, one of the weights
$W_{m,n}(s,\{Q^2_\ell\})$) and find a set of $n$, $\{ t_j\}$ and $\{ x_j\}$
such that the associated $w_n(E;\{t_j\},\{x_j\})$ represents a close
approximation to the weight $2E\, W(s=E^2)$ appearing in the
alternate, $E$-dependent expression
\begin{equation}
\label{moldspecint}
\int_{s_{\rm th}}^\infty  ds\, W(s) \r(s)\,
=\, \int_{E_{\rm th}}^\infty dE\, 2E\,  W(E^2)\, \r(E^2)
\end{equation}
for the $W(s)$-weighted spectral integral.
If such a choice exists, the associated sum rule, Eq.~(\ref{sumruleexponpre}),
will involve a spectral integral whose weighting, by construction, is
similar to that of the desired original $W(s)$-weighted spectral integral,
but whose right-hand side involves values of $C(t)$ at only the finite
number of chosen $t$.

The key point is that it is indeed possible, using the method of
Ref.~\cite{HLT}, to choose $n$, the $\{ t_j\}$ and the $\{ x_j\}$ such
that $w_n(E;\{t_j\},\{x_j\})$ provides a very good approximation to
$2E\, W(s=E^2)$ for weight choices, like $W(s)=W_{m,n}(s,\{Q^2_\ell\})$,
of interest in exploring the $a_\mu^{\rm HVP}$ problem. While in what
follows we focus, to be specific, on examples with
$W(s)=W_{m,n}(s=E^2;\{Q_\ell^2\})$, we stress that the method is
applicable to more general initial weight choices as well.

The construction of Ref.~\cite{HLT} proceeds as follows. Starting
with an initial desired weight $W(s)$, to be referred to in what
follows as the ``pre-mold,'' we choose a set of (positive) $t$ values
$\{ t_j\}$, $j=1,\cdots ,n$, and minimize
\begin{equation}
\label{min}
\int_{E_{\rm th}}^\infty
{\frac{dE}{E^4}}\,\biggl|w_n(E;\{t_j\},\{x_j\})-2E\, W(E^2)\biggr|^2
\end{equation}
with respect to the parameters $x_j$, $j=1,\dots,n$. We denote the
parameter values which accomplish this minimization by $\{ x_j^W\}$.
The result is a $w_n(E;\{t_j\},\{x^W_j\})$ which represents a (close)
approximation to $2E\, W(E^2)$, one having the form of
$E^2$ times a weighted sum of exponentials. We will refer to the product $2E\, W(E^2)$ as the ``mold,''
and the approximation $w_n(E;\{t_j\},\{x^W_j\})$ as the ``cast.''
The sum rule Eq.~(\ref{sumruleexponpre}) then takes the form
\begin{equation}
\label{sumruleexpon}
I_{W^\prime}\, \equiv\, \int_{E_{\rm th}}^\infty dE\,
\r(E^2)w_n(E;\{t_j\},\{x_j^W\}) =\sum_{j=1}^n x_j^W C(t_j)\ ,
\end{equation}
where the subscript/superscript $W$ emphasizes the role of the underlying
pre-mold function, $W(s)$, while the prime reminds us that the spectral
integral is evaluated with the derived cast $w_n$. We will refer to sum
rules of the form~(\ref{sumruleexpon}) as ``exponential-weight sum rules''
(EWSRs). Equation~(\ref{min}) is solved by
\begin{equation}
\label{sol}
x_i=\sum_{j=1}^n A^{-1}_{ij}f_j\ ,
\end{equation}
with the matrix $A$ and the input vector $f$ defined by
\begin{equation}
\label{defAf}
A_{ij}=\int_{E_{\rm th}}^\infty dE\,e^{-(t_i+t_j) E}\ ,\qquad
f_i=2\int_{E_{\rm th}}^\infty dE\,e^{-t_i E}W(E^2)/E\ .
\end{equation}
At this point, we depart from the philosophy of Ref.~\cite{HLT}: we throw
away the mold and keep the cast. If the cast is a good approximation
of the mold, it will serve equally well for comparing experimental data,
represented by $\r(s)$, with lattice data, represented by $C(t)$, in the
region of energies characterized by the mold, and there is thus no need
to keep the mold. Instead, we work directly with the cast sum rule,
Eq.~(\ref{sumruleexpon}).

Following this philosophy, there is also no need to keep the values of the
$x_j$ obtained from Eq.~(\ref{sol}) to a very high precision. One can keep
a fixed number of digits, and declare the values chopped off after
the last of these digits as the exact values of the $x_j$ to be used.
The only requirement is that the thus-defined exponential weight still
fullfills the goal for which it was designed, \ie, that it probes the
desired energy range. In our examples in the next section, we will chop
off the $x_j$ values at six digits, \seef\ Eqs.~(\ref{xs15}) and
~(\ref{xs25}).

The only role of the pre-mold and mold functions in this approach
is to fix the type of weighting one wants in the weighted spectral
integral in cases (such as exploring the BaBar--KLOE discrepancy) where
the desired weighting is naturally formulated in the $s$-space spectral
integral representation. The cast then provides a similarly weighted
spectral integral having the advantage that the corresponding weighted
Euclidean integral is trivially written down using the exact sum rule
~(\ref{sumruleexpon}) above. Attempting to find a set of $x_j$ which
produce such a desired spectral integral weighting is a much more
challenging task without the intermediate step of first setting up the
mold function and using it to find a set of $x_j$ which implement the
Euclidean representation corresponding to the closely related
$s$-dependent cast weighting. We will discuss examples of this
strategy in Sec.~\ref{exsumrule2}, where the pre-molds will be
the weights we consider in Sec.~\ref{exsumrule1}.

It is important to realize that, once one has a cast, {\it i.e.}, one has chosen the
set $\{ t_j\}$ and a strategy for determining the associated coefficients
$\{ x_j\}$, the resulting sum rule, (2.18), is \emph{exact}. Different choices
for the set $\{ t_j\}$ produce different cast functions, even when starting
from the same pre-mold. These different casts, moreover, all differ from
their original pre-mold(s). The exact cast sum rules corresponding to
different cast choices are then just that, different exact sum rules in
their own right, and not to be thought of as approximations to the (equally
exact) pre-mold sum rule.     Differences in the values of the left-hand-sides
    (or right-hand sides) of the sum rules (2.18) corresponding to different
    cast choices thus simply reflect the fact that those sum rules involve
    different weights. Such differences in no way constitute an additional
    systematic uncertainty on the values of any of the individual cast sum rules.{\footnote{This remark should be kept in mind when
    considering the results in Section III.B below. The results quoted in the
    first (or second) lines of Eqs.~(\ref{HLTRHS}) and (\ref{HLTRHShat}), for example, correspond
    to the same pre-mold choice, but different choices of the related cast. The
    (slightly different) cast choices produce (slightly different) central values
    for the (weight-dependent) right-hand sides of the corresponding cast sum
    rules. These values, of course, also differ slightly from those of the right-hand
    sides of the sum rules corresponding to the (also slightly different) pre-mold
    weights, given in Eqs.~(\ref{RHS}). The results of Eqs.~(\ref{RHS}), (\ref{HLTRHS}) and (\ref{HLTRHShat}) are
    those of three independent sum rules involving three different weights.}}

To illustrate the freedom to choose different casts, we discuss one more variant. For the types of
weight functions we consider in this paper, the matrix $A$ defined in
Eq.~(\ref{defAf}) may have very small eigenvalues, and, correspondingly, the
$x^W_j$ values may span a rather large range (as in Eqs.~(\ref{xs15}) and
~(\ref{xs25}) below), leading to potentially sizeable cancellations and
possible error inflation. We may reduce this range by modifying the
matrix $A$ to some extent, while keeping the cast close to the mold, so
that the modified sum rule still probes essentially the same
range in $\sqrt{s}$. The specific modification we will consider is to
replace the matrix $A$ with the related matrix $\hat{A}$, defined by
\begin{equation}
\label{Ahat}
\widehat{A}(\l)=(1-\l)A+\l {\bf 1}_n\ ,
\end{equation}
where ${\bf 1}_n$ is the $n\times n$ unit matrix. This simply replaces
eigenvalues of $A$ much smaller than $\l$ with new eigenvalues of
order $\l$, leaving eigenvalues much larger than $\l$ essentially
unchanged. We thus expect, and indeed find to be the case below,
that if we choose $\l$ larger than the smallest eigenvalue of $A$
the range of the $x^W_j$ values, and hence the error on the lattice
sides of the resulting EWSRs, will be reduced. We will denote the
analogue of the spectral integral $I_{W^\prime}$ of Eq.~(\ref{sumruleexpon})
obtained using the modified set of $x^W_j$ resulting from this
replacement by $I_{\widehat{W}}$ in what follows.

    We close this subsection with a recap of the key points of the motivation
for the EWSR construction. The goal of this paper is to provide a method
for exploring the tension between lattice and dispersive results for
$a_\mu^{\rm HVP}$ by identifying $s$-dependent weights, $W(s)$, which are
simultaneously localized to relatively narrow regions in $s$ and such that
the alternate, Euclidean-time ($t$) representations of the $W(s)$-weighted
spectral integrals, $I_W$, evaluated using input lattice data, have errors
small enough to make the resulting dispersive-lattice comparison numerically
interesting. Such dispersive-lattice comparisons are, in principle, possible
for any $W(s)$, provided both the spectral data and lattice data are known
to sufficiently high precision. In practice, however, lattice errors grow
rapidly at large Euclidean $t$. The $I_W$ produced by the vast majority of
$W(s)$ have, as in the case of the RWSR examples above, equivalent
Euclidean-$t$ representations which contain contributions from $C(t)$ at
arbitrarily large $t$, and hence typically have enhanced errors when
evaluated using real-world lattice data. The EWSR construction, in
restricting by hand the range of Euclidean $t$ for which lattice results
for $C(t)$ are required, allows one to mitigate this problem, while at the
same time retaining the desired qualitative $s$-dependent weighting.

     Another way of describing what is going on in the EWSR construction is
as follows: given a typical pre-mold, $W(s)$, chosen with a particular
localization in $s$ in mind, the Euclidean-$t$ representations equivalent
to the spectral integrals produced by, not just that pre-mold, but also
the vast majority of qualitatively similar ``nearby'' weights, will all
suffer from the error-enhancing feature of having non-trivial support at
large Euclidean $t$. The EWSR construction, however, shows that this is
not a generic feature of all such ``nearby'' weights: weights $V(s)$
qualitatively similar to $W(s)$ exist for which the corresponding
weighted spectral integral has a Euclidean-$t$ representation with
no support whatsoever beyond the maximum $t$ employed in constructing
$V(s)$. We expect such special ``nearby'' weights to produce equivalent
lattice evaluations of $I_V$ with significantly reduced relative errors,
and show, in Section III.B, that this expectation is, indeed, borne out.
An important lesson to be drawn from this observation is that, when
real-world lattice errors are taken into account, $s$-dependent weights
which are very similar as functions of $s$, may differ significantly in how
accurately the equivalent lattice representations of their weighted spectral
integrals can be evaluated. Sum rules based on judicious choices of EWSR
weights, $V(s)$, constructed starting from generic pre-molds, $W(s)$,
are thus likely to represent better choices for use in exploring
dispersive-lattice differences than would the apparently rather similar
sum rules based on the underlying generic pre-molds.

\section{\label{comparison} Example: application to comparison between experimental and lattice data}

\begin{figure}
\vspace*{4ex}
\begin{center}
\includegraphics*[width=10cm]{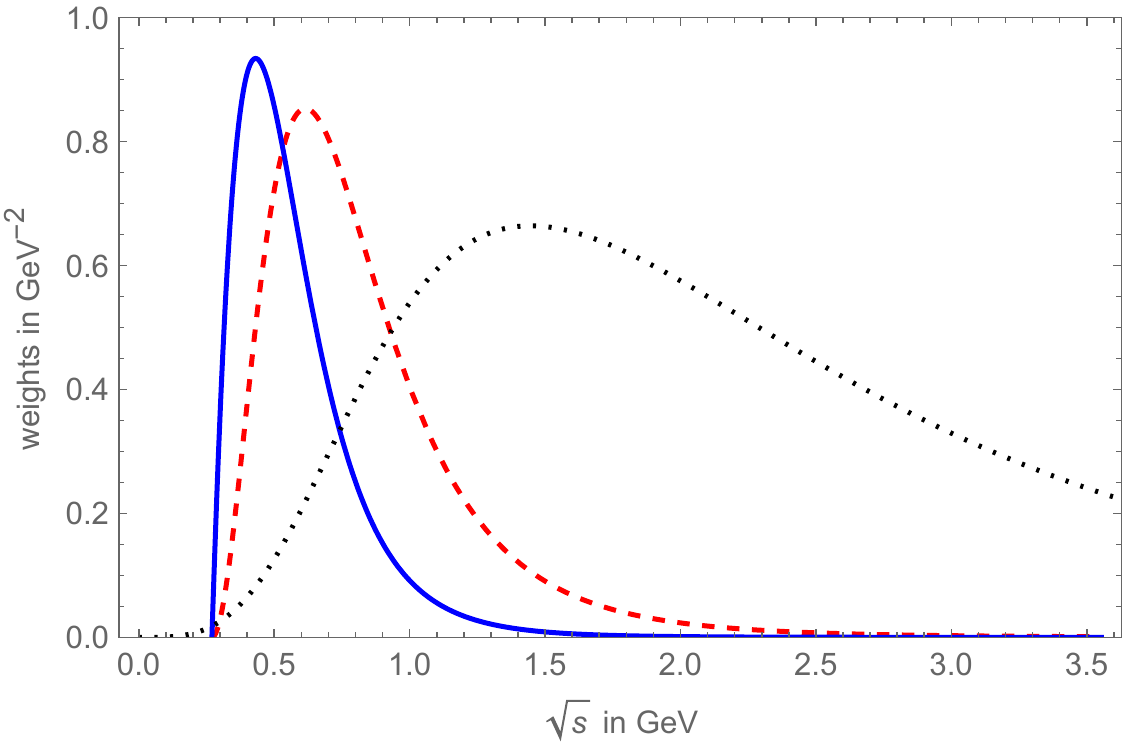}
\end{center}
\begin{quotation}
\floatcaption{weights}%
{{\it The weights $W_{1,5}$ (blue, solid), and $15\times W_{2,5}$
(red, dashed) with $Q^2_i$ defined in Eq.~(\ref{Qs15}). ($W_{2,5}$
is multiplied by a factor 15 for clarity.) The black dotted curve
shows the ``intermediate window'' weight of Ref.~\cite{RBC},
transformed to the equivalent $\sqrt{s}$-dependent form \cite{Rwindow}.}}
\end{quotation}
\vspace*{-4ex}
\end{figure}
We will now turn to two numerical examples, using two differents weights
$W_{m,n}$ of the type~(\ref{srweight}), both with $s_{\rm th}=4m_\p^2$
and
\begin{equation}
\label{Qs15}
Q_\ell^2=0.25+0.075(\ell-1)\ \mbox{GeV}^2\ ,\qquad \ell=1,\dots,5\ .
\end{equation}
The first has $m=1$, $n=5$, and the second $m=2$, $n=5$.
We choose $m_\p=134.977$~MeV and $m_\t=1776.86$~MeV. These two weights
are plotted as a function of $\sqrt{s}$ in Fig.~\ref{weights}, where
$W_{2,5}$ has been multiplied by a factor 15 to show it on approximately
the same scale as $W_{1,5}$.\footnote{The absolute scale of the weights
is arbitrary, as long as the same scale is used on both sides of the sum
rules.} The weight $W_{2,5}$ overlaps better with the region
dominated by the $\r$, but falls off more slowly for large $s$. The
value of $n-m$ is smaller for $W_{2,5}$ than for $W_{1,5}$, which leads
us to expect a smaller error on the right-hand side of the $W_{2,5}$ sum
rule, according to App.~\ref{precision}. For comparison, we also show
the Minkowski version \cite{Rwindow} of the ``intermediate window''
weight introduced as a function of $t$ in Ref.~\cite{RBC}. These
weights will be used for a study of the sum rule~(\ref{sumruleCt}) in
Sec.~\ref{exsumrule1} and a study of the sum rule~(\ref{sumruleexpon})
in Sec.~\ref{exsumrule2}.

\begin{figure}
\vspace*{4ex}
\begin{center}
\includegraphics*[width=10cm]{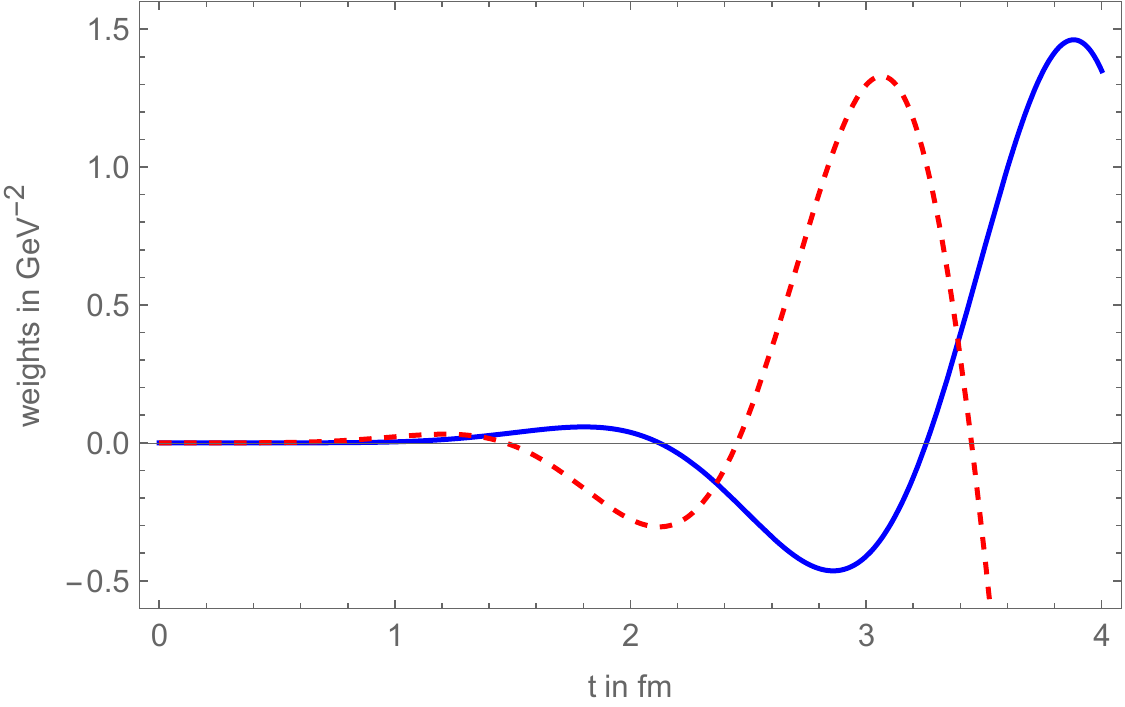}
\end{center}
\begin{quotation}
\floatcaption{cmn}%
{{\it The weights $10^{-5}c^{(1,5)}(t)$ (blue), and
$15\times 10^{-5}c^{(2,5)}(t)$ (red) with $Q^2_\ell$ as in Eq.~(\ref{Qs15}).
}}
\end{quotation}
\vspace*{-4ex}
\end{figure}

In Fig.~\ref{cmn} we show the functions $c^{(1,5)}(t)$ and $c^{(2,5)}(t)$
defined in Eq.~(\ref{cmndef}), multiplied by $10^{-5}$ and $15\times 10^{-5}$,
respectively. We see that these functions have the bulk of their support
toward larger $t$, while being very small at values of $t$ smaller than
approximately 1~fm. This is not surprising, as the weights $W_{1,5}(s)$
and $W_{2,5}(s)$ have their support at relatively small $\sqrt{s}$. We
will see below to what extent this has an effect on the size of the errors
with which the right-hand side of Eq.~(\ref{sumruleCt}) can be evaluated.

In these examples, we will use the combined $R$-ratio data from
Ref.~\cite{KNT19} to evaluate the weighted spectral integrals on the
left-hand side (LHS) and the light-quark-connected lattice data from
Ref.~\cite{ABGP22} to evaluate the weighted integrals over $C(t)$ on the
right-hand side (RHS).{\footnote{{The results of Ref.~\cite{ABGP22} were obtained using
ensembles from Refs.~\cite{MILC,CalLat}.}}} Since the data from Ref.~\cite{ABGP22} are for the
light-quark-connected part only, this is a comparison between apples
and oranges as far as the central results are concerned. Here, however,
we are interested in seeing the size of typical errors on each side of
the sum rule, and using these immediately accessible recent data sources
suffices for this purpose. We emphasize that our intent is to
investigate the size of lattice and spectral integral errors in our
proposed methodology, and that, at present, no conclusions should be drawn
from the level of numerical agreement between the two sides of the
sum rules considered throughout Sec.~\ref{comparison}. We also neglect
corrections to the lattice data of Ref.~\cite{ABGP22} for finite-volume,
taste-breaking and pion-mass mistuning effects, having
convinced ourselves that these corrections are small, and do not lead to
qualitative changes in the conclusions we obtain below. This is
consistent with our intent to study the methodology of these
sum rules, leaving concrete applications to the future.

\subsection{\label{exsumrule1} Examples using rational-weight sum rules}
We begin with the values of the spectral integrals appearing
on the LHS of the sum rule~(\ref{sumruleCt}) for the
weights $W_{1,5}$ and $W_{2,5}$ obtained using the $R$-ratio data
from Ref.~\cite{KNT19}. We find
\begin{eqnarray}
\label{LHS}
I_{W_{1,5}}({\rm LHS})&=&0.4756(16)\ ,\\
I_{W_{2,5}}({\rm LHS})&=&0.09107(34)\ .\nonumber
\end{eqnarray}
Both spectral integrals are obtained with a precision better than $0.4\%$.

\begin{table}[t]
\begin{center}
\begin{tabular}{|c|c|c|c||c|c|}
\hline
label & $a$ (fm) & $L^3\times T$ & $m_\p$ (MeV) & $I_{W_{1,5}}({\rm RHS})$
&  $I_{W_{2,5}}({\rm RHS})$\\
\hline
96 & 0.05684 & $96^3\times 192$ & 134.3 & 0.463(45) & 0.0860(74) \\
64 & 0.08787 & $64^3\times 96$ & 129.5 & 0.453(23) & 0.0826(28) \\
48I & 0.12121 & $48^3\times 64$ & 132.7 & 0.426(13) & 0.0831(13) \\
32 & 0.15148 & $32^3\times 48$ & 133.0 & 0.433(13) & 0.0801(17) \\
48II & 0.15099 & $48^3\times 64$ & 134.3 & 0.411(14) & 0.0824(18) \\
\hline
\end{tabular}
\end{center}
\vspace*{-3ex}
\begin{quotation}
\floatcaption{tab:ensembles}{{\it The right-hand sides
of Eq.~(\ref{sumruleCt}) for the lattice ensembles of Ref.~\cite{ABGP22}.
The columns to the left of the double vertical line contain
the label of the ensemble, the lattice spacing $a$, the
spatial/temporal extent, $L^3\times T$ (in lattice
units), and the Nambu--Goldstone pion mass $m_\p$. The columns
to the right of the double line contain the integrals
$I_{W_{1,5}}({\rm RHS})$ and $I_{W_{2,5}}({\rm RHS})$. }}
\end{quotation}
\vspace*{-4.5ex}
\end{table}
We next turn to the evaluation of the RHSs of the $W_{1,5}$ and
$W_{2,5}$ RWSRs using lattice data for the light-quark connected
contribution to the correlator $C(t)$ obtained using the five ensembles
of Ref.~\cite{ABGP22}. Table~\ref{tab:ensembles} shows the ensemble parameters
to the left of the double vertical line, and the associated
finite-lattice-spacing results for $I_{W_{1,5}}({\rm RHS})$ and
$I_{W_{2,5}}({\rm RHS})$ to the right of it. We
note that the values of the pion mass and the volume differ
from ensemble to ensemble, and would, in principle, need to be adjusted to
common values before carrying out an extrapolation to the continuum limit.
Since, however, the current statistical errors are large enough that these
mistunings would make no difference in practice, we have, instead,
neglected such corrections, and carried out an extrapolation
linear in $a^2$ to the continuum limit. For the $a=0.15$~fm lattice spacing
we chose the larger of the two volumes, not using the 32 ensemble. The fits
have excellent $p$-values, and we find the continuum-limit values
\begin{eqnarray}
\label{RHS}
I_{W_{1,5}}({\rm RHS})&=&0.468(26)\ ,\\
I_{W_{2,5}}({\rm RHS})&=&0.0838(33)\ .\nonumber
\end{eqnarray}
These integrals have a precision of $5.6\%$, respectively,
$3.9\%$.\footnote{The relatively large errors make the figures
showing these fits not very interesting, and we thus omit them.
In the next subsection we will show the equivalent fits for the
EWSRs, for which the lattice values have smaller errors.} Based
on these results we make the following observations. First, the errors
in Eq.~(\ref{RHS}) are large, and a significant reduction in errors
would be needed to make these sum rules useful. However, as
we will see in the next subsection, errors on the RHS, obtained
using the same lattice data, are smaller for sum rules of the
type~(\ref{sumruleexpon}) than for those of the type~(\ref{sumruleCt}).
Second, from the different
size of the errors for the sum rules with $W_{1,5}$ and $W_{2,5}$, it is
clear that a fine tuning of the rational weight may be necessary to make
optimal use of a certain set of lattice data. We emphasize again that the
numbers in Eqs.~(\ref{LHS}) and~(\ref{RHS}) should not be directly compared in
this example, because, in this pilot study, we used the fully inclusive
data on the LHS, but only light-quark-connected data on the
RHS. Of course, fully inclusive lattice data are becoming available,
and with such data, a direct comparison will be possible. It is,
moreover, reasonable to expect such lattice data to have significantly
improved errors, allowing for significant improvements to the relative
errors in Eq.~(\ref{RHS}) as well.

\subsection{\label{exsumrule2} Examples using exponential-weight sum rules}
We now turn to sum rules of the type~(\ref{sumruleexpon}), using as
examples of the pre-mold the weights $W_{1,5}$ and $W_{2,5}$ of
Sec.~\ref{exsumrule1}. We still need to choose the $t_j$ appearing in
Eq.~(\ref{sumruleexpon}) and, in this section, will employ the following
values, given in units of GeV$^{-1}$:
\begin{equation}
\label{ts}
t_1=3\ ,\quad t_2=6\ ,\quad t_3=9\ ,\quad t_4=12\ ,\quad
t_5=15
\ .
\end{equation}
Since these values do not exactly coincide with the values of $t$ available
for the lattice ensembles we employ, we obtain the associated $C(t_j)$
values, in all cases, by linear interpolation from the nearest available
$t$, taking into account all correlations in the interpolation. The range of
values in Eq.~(\ref{ts}) is one for which lattice computations of $C(t)$
generally have small statistical errors.

With the $t_j$ defined, the values of the coefficients $x_j$ are fixed by
the pre-mold choice. For $W'_{1,5}$ we find
\begin{equation}
\label{xs15}
W'_{1,5}:\quad
x_1=37.4123\ ,\ x_2=2625.13\ ,\ x_3=25912.1\ ,\ x_4=-106707.\ ,\ x_5=78192.8\ ,
\end{equation}
while for $W'_{2,5}$ we find
\begin{equation}
\label{xs25}
W'_{2,5}:\quad
x_1=34.0249\ ,\ x_2=870.640\ ,\ x_3=-5501.14\ ,\ x_4=9933.01\ ,\ x_5=-5284.24
\ .
\end{equation}
For $\widehat{W}_{1,5}$ and $\widehat{W}_{2,5}$, obtained by replacing $A$
by the $\l=10^{-9}$ version of $\widehat{A}$ of Eq.~(\ref{Ahat}) in
Eq.~(\ref{sol}), we find
\begin{equation}
\label{xhats15}
\widehat{W}_{1,5}:\quad
x_1=-78.8487\ ,\ x_2=5688.30\ ,\ x_3=2223.96\ ,\ x_4=-36638.0\ ,\
x_5=8047.38\ ,
\end{equation}
and
\begin{equation}
\label{xhats25}
\widehat{W}_{2,5}:\quad
x_1=44.8916\ ,\ x_2=590.933\ ,\ x_3=-3373.53\ ,\ x_4=3716.86\ ,\ x_5=879.149\ .
\end{equation}

As already explained in Sec.~\ref{sumrule2}, we will use these exact values in
the rest of this subsection.

In the upper panels of Fig.~\ref{castmold} we show the molds
$2\sqrt{s}\, W_{1,5}(s)$ and $2\sqrt{s}\, W_{2,5}(s)$ together with their
associated casts, $W_{1,5}^\prime (s)=w_5(E;\{t_j\},\{x_j^{W'_{1,5}}\})$
and $W_{2,5}^\prime (s)=w_5(E;\{t_j\},\{x_j^{W'_{2,5}}\})$, determined
by Eqs.~(\ref{ts}) and~(\ref{xs15}),~(\ref{xs25}). We also show the casts
$\widehat{W}_{1,5}(s)=w_5(E;\{t_j\},\{x_j^{\widehat{W}_{1,5}}\})$ and
$\widehat{W}_{2,5}(s)=w_5(E;\{t_j\},\{x_j^{\widehat{W}_{2,5}}\})$,
determined with Eqs.~(\ref{xhats15}) and~(\ref{xhats25}) instead.
On the scale of these figures,
the casts are almost indistinguishable from the underlying molds for the
$A$-based casts, and quite close for the $\widehat{A}$-based casts.
The smaller-vertical-scale lower panels show the corresponding cast--mold
differences in more visually evident form.

\begin{figure}
\vspace*{4ex}
\begin{center}
\includegraphics*[width=7cm]{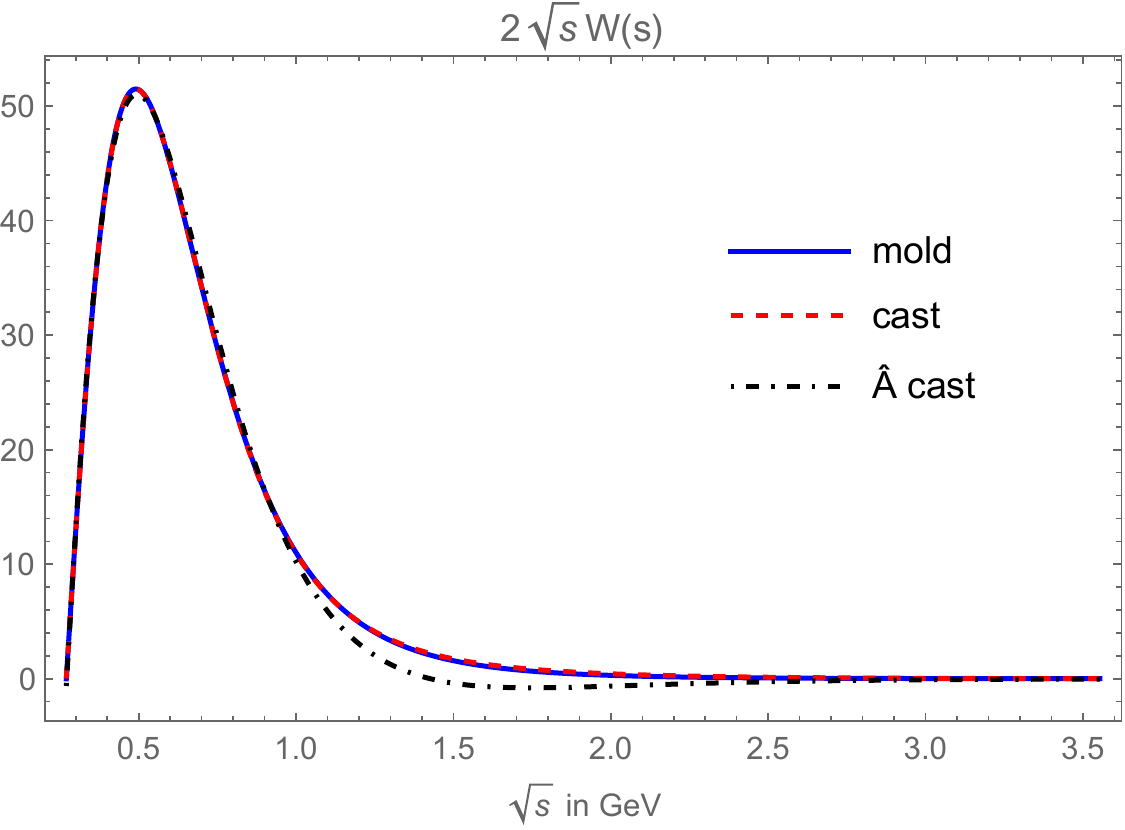}
\hspace{2ex}
\includegraphics*[width=7cm]{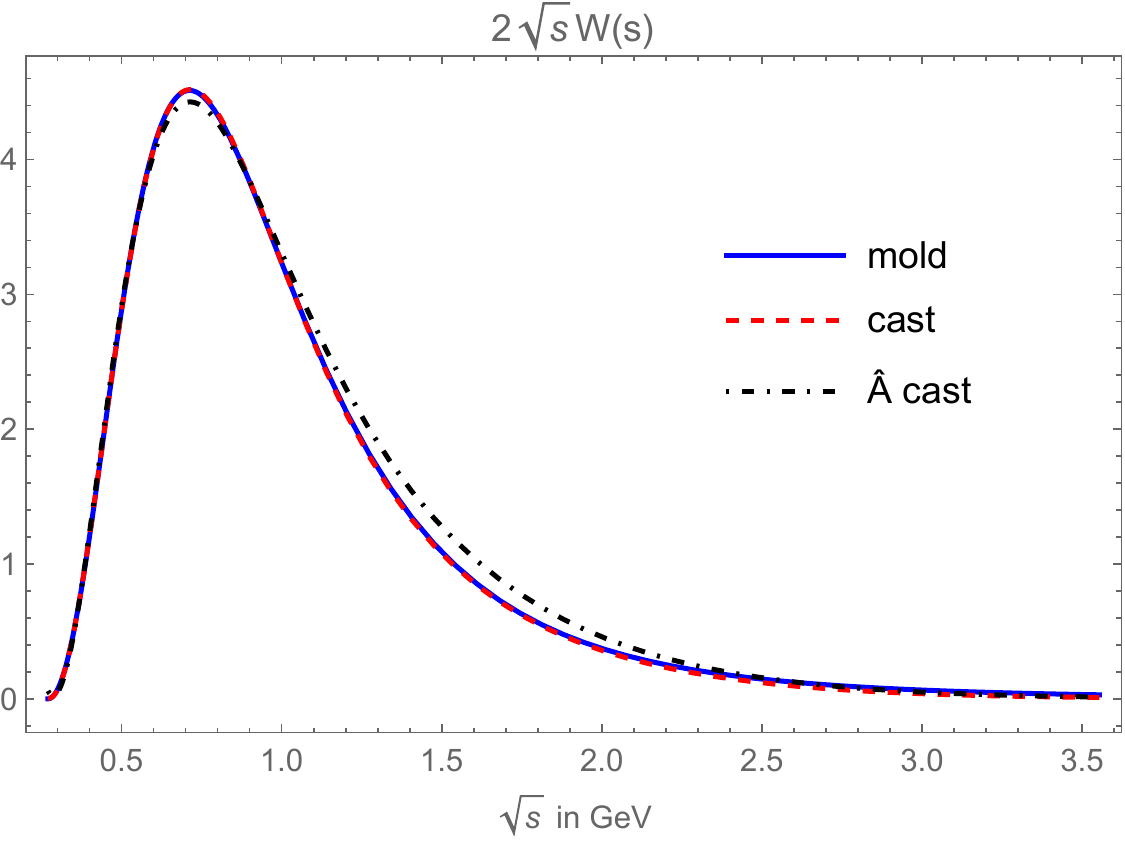}
\vspace*{4ex}
\includegraphics*[width=7cm]{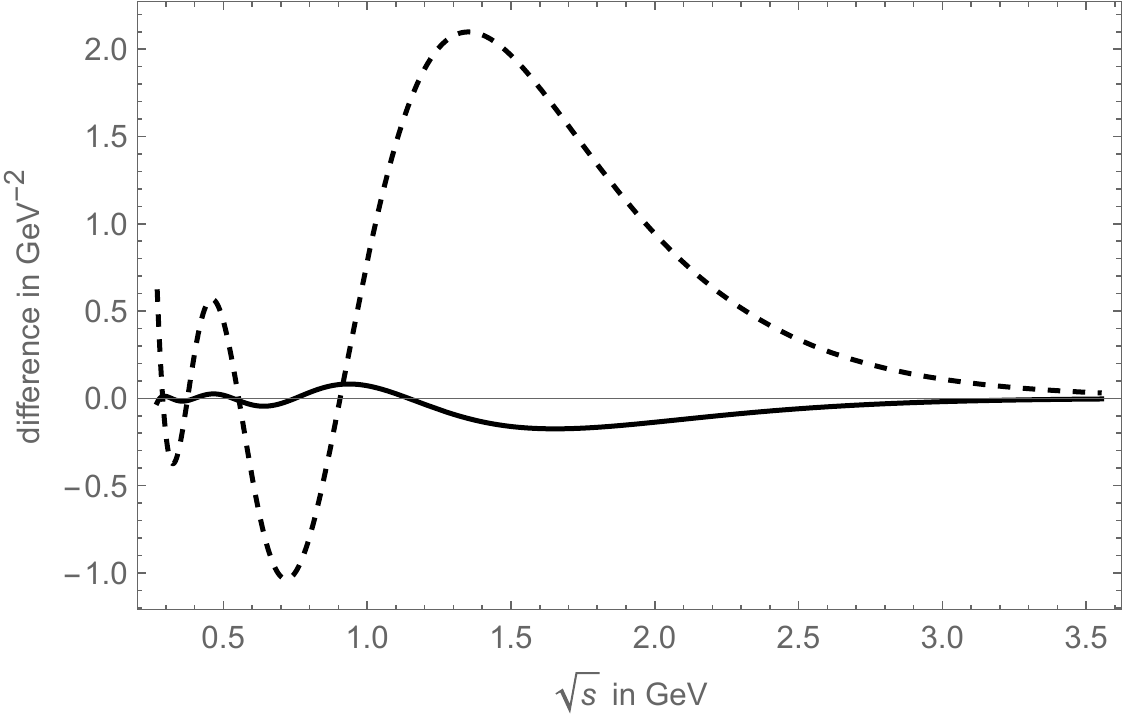}
\hspace{2ex}
\includegraphics*[width=7cm]{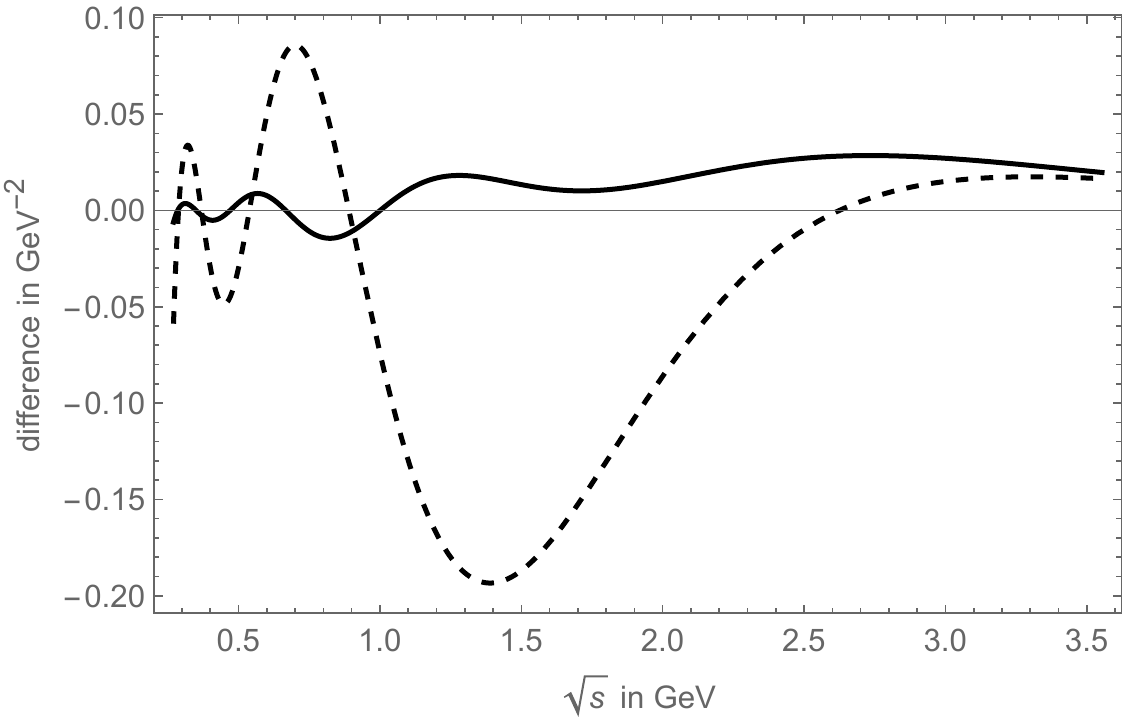}
\end{center}
\vspace*{-8ex}
\begin{quotation}
\floatcaption{castmold}%
{{\it Upper panels: molds (blue continuous curves) and casts (red dashed
curves) for the molds $2\sqrt{s}\,W_{1,5}(s)$ (left panel) and
$2\sqrt{s}\,W_{2,5}(s)$ (right panel). The black dot-dashed curves show
the casts obtained with $\widehat{A}$ of Eq.~(\ref{Ahat}) instead of $A$ of
Eq.~(\ref{defAf}), choosing $\l=10^{-9}$.
Lower panels: differences between mold and cast, for each case (with the
dashed curves the difference for the $\widehat{A}$-based casts). Note the
very different scales on the vertical axes in the upper and lower
figures.}}
\end{quotation}
\vspace*{-4ex}
\end{figure}

As for the RWSR cases, we first give the LHS values,
$I_{W^\prime_{1,5}}$ and $I_{W^\prime_{2,5}}$, obtained by
integrating over the $R$-ratio data:
\begin{eqnarray}
\label{HLTLHS}
I_{W_{1,5}^\prime}({\rm LHS})&=&0.4788(16)\ ,\\
I_{W_{2,5}^\prime}({\rm LHS})&=&0.08922(34)\ .\nonumber
\end{eqnarray}
For $I_{\widehat{W}_{1,5}}$ and $I_{\widehat{W}_{{2,5}}}$ we find
\begin{eqnarray}
\label{HLTLHShat}
I_{\widehat{W}_{1,5}}({\rm LHS})&=&0.4488(16) \ ,\\
I_{\widehat{W}_{2,5}}({\rm LHS})&=&0.09214(37)\ .\nonumber
\end{eqnarray}
As before, the $R$-ratio data of Ref.~\cite{KNT19} allow us to determine
these values with errors $\ltap\, 0.4\%$. We note that the values
in Eq.~(\ref{HLTLHS}) or Eq.~(\ref{HLTLHShat}) are not the same as those
in Eq.~(\ref{LHS}), even if they are rather close. This reflects the fact
that mold and cast weights are not identical.

Because of the well-known even-odd time oscillation of staggered correlators,
we define smoothed-out correlators
\begin{equation}
\label{smooth}
\widehat{C}(t)=\frac{1}{4}\left(C(t-1)+2C(t)+C(t+1)\right)\ .\\
\end{equation}
We will use the averaged correlator $\widehat{C}(t)$ when evaluating the
RHS of Eq.~(\ref{sumruleexpon}).\footnote{More sophisticated
approaches are possible, but we consider this average sufficient for
the explorative nature of this section.}

We show the results for $I_{W_{1,5}^\prime}({\rm RHS})$,
$I_{W_{2,5}^\prime}({\rm RHS})$, $I_{\widehat{W}_{1,5}}({\rm RHS})$ and
$I_{\widehat{W}_{2,5}}({\rm RHS})$ in Table~\ref{tab:HLT}. Using the
values obtained with the three smallest lattice spacings,
\ie, the ensembles 96, 64 and 48I, and performing fits linear
in $a^2$, we find the continuum-limit values
\begin{eqnarray}
\label{HLTRHS}
I_{W_{1,5}^\prime}({\rm RHS})&=&0.496(17)\ ,\\
I_{W_{2,5}^\prime}({\rm RHS})&=&0.0798(18)\ ,\nonumber
\end{eqnarray}
and
\begin{eqnarray}
\label{HLTRHShat}
I_{\widehat{W}_{1,5}}({\rm RHS})&=&0.4669(68)\ ,\\
I_{\widehat{W}_{2,5}}({\rm RHS})&=&0.0824(10)\ .\nonumber
\end{eqnarray}
These linear fit results are shown by the red lines in
Fig.~\ref{fitsEWSR}. The blue curves in the same figure show
the results of alternate fits quadratic in $a^2$ obtained using the
four ensembles 96, 64, 48I and 48II. The linear and quadratic
fits give consistent continuum-limit results.

\begin{table}[t]
\begin{center}
\begin{tabular}{|c||c|c||c|c|}
\hline
label &   $I_{W_{1,5}^\prime}({\rm RHS})$
&   $I_{W_{2,5}^\prime}({\rm RHS})$
&   $I_{\widehat{W}_{1,5}}({\rm RHS})$
&   $I_{\widehat{W}_{2,5}}({\rm RHS})$\\
\hline
96  & 0.504(20) & 0.0799(23) & 0.4625(75) & 0.0838(12) \\
64 & 0.4822(92) & 0.08703(94) & 0.4520(38) & 0.08994(54) \\
48I & 0.4790(72) & 0.09187(80) & 0.4401(40) & 0.09558(51) \\
\hline
32 & 0.498(12) & 0.10073(76) & 0.4398(68) & 0.10623(45) \\
48II & 0.4837(85) & 0.1009(11) & 0.4301(46) & 0.10605(64) \\
\hline
\hline
$a=0$ & 0.496(17) & 0.0798(18) & 0.4669(68) & 0.0824(10)  \\
\hline
$p$-value & 0.44 & 0.15 & 0.78 & 0.10 \\
\hline
rel. error & 3.4\% & 2.2\% & 1.5\% & 1.2\% \\
\hline
\end{tabular}
\end{center}
\vspace*{-3ex}
\begin{quotation}
\floatcaption{tab:HLT}{{\it The right-hand side of Eq.~(\ref{sumruleexpon})
for the lattice ensembles of Ref.~\cite{ABGP22}. The 2nd and 3rd columns
give, respectively, the values for $I_{W_{1,5}^\prime}$ and
$I_{W_{2,5}^\prime}$, obtained using Eq.~(\ref{smooth}). The 4th and 5th columns
give, respectively, the values for $I_{\widehat{W}_{1,5}}$ and
$I_{\widehat{W}_{2,5}}$, obtained using $\widehat{A}$ of
Eq.~(\ref{Ahat}) instead of $A$, with $\l=10^{-9}$.
Below the double horizontal line we show (i) the continuum
extrapolation for each column, obtained using fits linear in $a^2$
to the values for the first three ensembles (96, 64 and 48I), (ii) the
$p$-value of that fit, and (iii) the relative statistical error of the
continuum limit value.}}
\end{quotation}
\vspace*{-4.5ex}
\end{table}

As shown in the Table~\ref{tab:HLT}, the integrals,
$I_{W_{1,5}^\prime}({\rm RHS})$ and $I_{W_{2,5}^\prime}({\rm RHS})$
in Eqs.~(\ref{HLTRHS}) have precisions of $3.4\%$ and $2.2\%$, respectively, while
$I_{\widehat{W}_{1,5}}({\rm RHS})$ and $I_{\widehat{W}_{2,5}}({\rm RHS})$
in Eqs.~(\ref{HLTRHShat}) have precisions of $1.5\%$ and $1.2\%$, respectively.
We see that the modification~(\ref{Ahat}) indeed pays off. We
emphasize again that the LHS and RHS central values should not be
directly compared in this study. What is of interest here is the size of the
statistical errors in Eqs.~(\ref{HLTLHS}),~(\ref{HLTLHShat}),~(\ref{HLTRHS})
and~(\ref{HLTRHShat}). While those in Eqs.~(\ref{HLTLHS}) and~(\ref{HLTLHShat})
are the same as those in Eq.~(\ref{LHS}), those in Eqs.~(\ref{HLTRHS}) and
~(\ref{HLTRHShat}) are significantly smaller than those in Eq.~(\ref{RHS}). This
reflects the freedom to avoid, through the restricted set of {$t_j$ values
employed in Eqs.~(\ref{HLTRHS}) and~(\ref{HLTRHShat}), higher-error, large-$t$
contributions, and provides a concrete example of the potential enhanced
utility of EWSRs over their closely related RWSR counterparts.

The improvement represented by the EWSR construction is, of course,
expected to vary depending on the choice of the $\{  t_j\}$ and the
strategy chosen for obtaining the associated coefficients $\{  x_j\}$.
We have already seen that, for the set of $\{  t_j\}$ chosen above, using
the strategy of Eq. (2.21) produces generally smaller $\{  x_j\}$ and hence,
as a result of reduced cancellation, improved errors on the lattice sides
of the associated EWSRs. We thus expect other strategies for constructing
EWSR weights which limit the size of the $\{  x_j\}$ to also produce EWSRs
with reduced lattice-side relative errors. Such considerations suggest,
e.g., avoiding sets $\{  t_j\}$ in which the $t_j$ are too closely spaced
since reducing the spacing between adjacent $t_j$, in reducing the
difference between the associated basis functions $E^2\, \exp (-t_j E)$,
is likely to force an increase in the size of the associated coefficients
$\{  x_j\}$.{\footnote{An example of the impact of choosing an overly finely spaced set
$\{t_j\}$ is provided by EWSRs based on alternate versions,
$W^{\prime\prime}_{15}$ and $W^{\prime\prime}_{25}$, of the primed
weights above, obtained using the more finely spaced set, $\{t_j\} = \{3,\, 5,\, 7,\, 9,\, 11,\, 13,\, 15\}$~GeV$^{-1}$,
covering the same range in $t$ as
the set (\ref{ts}). This finer spacing produces relative errors on the
lattice sides of $W^{\prime\prime}_{15}$ and $W^{\prime\prime}_{25}$
EWSRs of 3.5\% and 3.3\%, respectively, which represents a 50\% error
increase in the case of $W^{\prime\prime}_{25}$ when compared with the
results of Table~\ref{tab:HLT}.}

\begin{figure}
\vspace*{4ex}
\begin{center}
\includegraphics*[width=7cm]{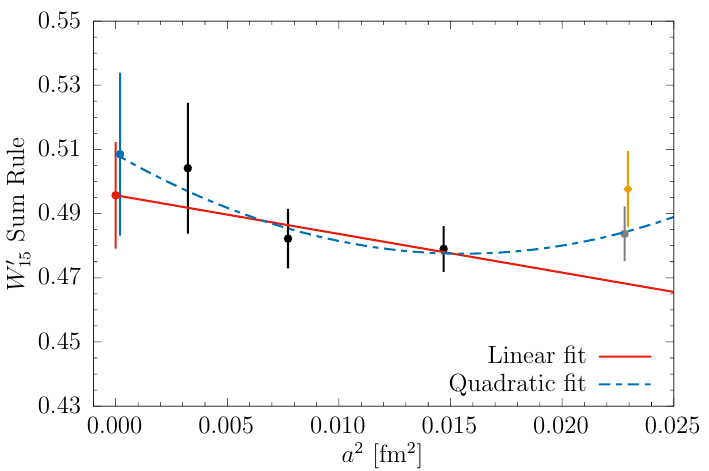}
\hspace{2ex}
\includegraphics*[width=7cm]{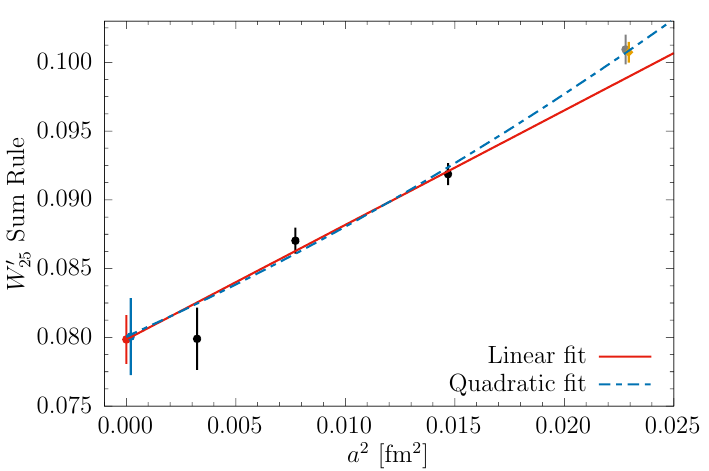}
\vspace{1ex}
\includegraphics*[width=7cm]{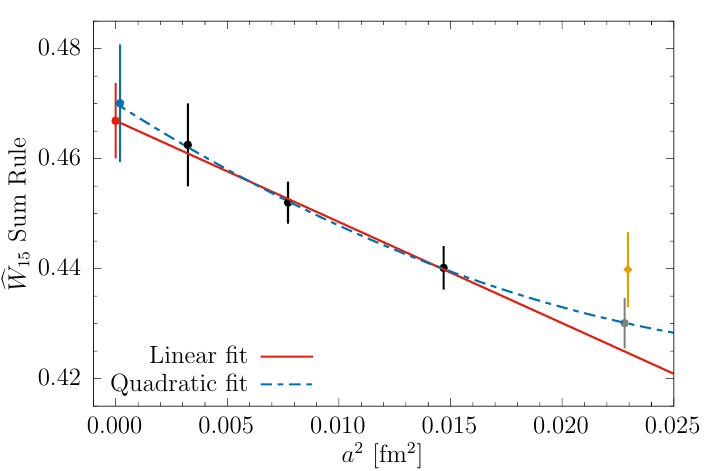}
\hspace{2ex}
\includegraphics*[width=7cm]{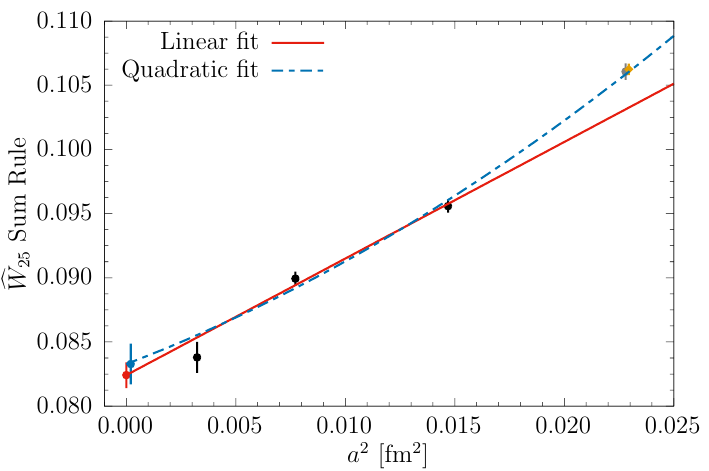}
\end{center}
\vspace*{-4ex}
\begin{quotation}
\floatcaption{fitsEWSR}%
{{\it Continuum extrapolations of $I_{W_{1,5}^\prime}({\rm RHS})$
(left upper panel) and $I_{W_{2,5}^\prime}({\rm RHS})$ (right upper panel),
and $I_{\widehat{W}_{1,5}}({\rm RHS})$ (left lower panel) and
$I_{\widehat{W}_{2,5}}({\rm RHS})$ (right lower panel).
The data points show the lattice results for the various ensembles.
Linear fits to the three left-most points (omitting the gray
filled circles) are shown by the red lines, and quadratic fits to all
four data points (the black and gray filled circles) by the dashed blue
curves. Results for the 32 ensemble, shown as orange diamonds,
are not used in either of these extrapolations.
}}

\end{quotation}
\vspace*{-4ex}
\end{figure}

We add several comments. First, it is likely that lattice data in the near
future will have smaller statistical errors than does the lattice
data of Ref.~\cite{ABGP22}, used in the current study. This makes it not
unlikely that sub-percent precision can be reached with EWSRs. Second,
we did not discuss systematic errors, which of course also need to be
controlled. In particular, for staggered fermions, smaller lattice
spacings than the current smallest value of about $0.06$~fm will
be needed \cite{ABGP22}. We thus do not claim that our linear fits to the
results in Table~\ref{tab:HLT} are the final word. However, similar fits to
improved data at smaller lattice spacings should allow these sum rules
to become of practical use. Finally, we note that the $W_{1,5}$
versions of Eqs.~(\ref{HLTLHS}) and~(\ref{HLTRHS}) are closer together than
the $W_{2,5}$ versions. This can be qualitatively understood. The
large-$t$ part of the full HVP correlator $C(t)$ with $j_\m=j'_\m$ the
hadronic electromagnetic current is dominated by the light-quark-connected
part, more so than the smaller-$t$ part. As can be seen from
Fig.~\ref{weights}, the weight $W_{1,5}$ has support at smaller $s$.
This translates into having support at larger $t$, as can be seen in
Fig.~\ref{cmn}. This explains the closer agreement for the $W_{1,5}$-based
case.  A similar observation also holds for Eqs.~(\ref{LHS}) and~(\ref{RHS}) in
Sec.~\ref{exsumrule1}.

\subsection{\label{BaBarKLOE} Comparison with BaBar--KLOE discrepancy}
One would like to get an idea about what is needed to make the lattice
errors in Eq.~(\ref{HLTRHS}) small enough for concrete applications of these
sum rules. To this end, we consider, as an example, the difference of the
values of $I_{W'_{1,5}}$ and $I_{W'_{2,5}}$ computed using the
BaBar \cite{BaBar,BaBar:2012bdw} or KLOE \cite{KLOE} versions of the
two-pion contribution to the EM spectral function. As Ref.~\cite{KLOE}
provides the $e^+e^-\rightarrow \pi^+\pi^-$ cross sections only up to
$s=0.95$~GeV$^2$, we will assume that the full discrepancy originates
from the difference in the measured cross sections below $s=0.95$~GeV$^2$.
This allows us to evaluate the two-pion contributions to
$I_{W_{1/2,5}}$, $I_{W'_{1/2,5}}$ and $I_{\widehat{W}_{1/2,5}}$
on both the BaBar and KLOE data. We find
\begin{eqnarray}
\label{BaBarKLOEdiff}
I_{W_{1,5}}^{\pi\pi}(\mbox{BaBar})
-I_{W_{1,5}}^{\pi\pi}(\mbox{KLOE})&=&0.0091(39)\ ,\\
I_{W_{2,5}}^{\pi\pi}(\mbox{BaBar})
-I_{W_{2,5}}^{\pi\pi}(\mbox{KLOE})&=&0.00152(51)\ ,
\nonumber \\
I_{W'_{1,5}}^{\pi\pi}(\mbox{BaBar})
-I_{W'_{1,5}}^{\pi\pi}(\mbox{KLOE})&=&0.0091(39)\ ,
\nonumber\\
I_{W'_{2,5}}^{\pi\pi}(\mbox{BaBar})
-I_{W'_{2,5}}^{\pi\pi}(\mbox{KLOE})&=&0.00153(52)\ ,
\nonumber \\
I_{\widehat{W}_{1,5}}^{\pi\pi}(\mbox{BaBar})
-I_{\widehat{W}_{1,5}}^{\pi\pi}(\mbox{KLOE})&=&
0.0094(40)\ ,\nonumber \\
I_{\widehat{W}_{2,5}}^{\pi\pi}(\mbox{BaBar})
-I_{\widehat{W}_{2,5}}^{\pi\pi}(\mbox{KLOE})&=&
0.00150(51)\ .\nonumber
\end{eqnarray}
Comparing these values with the lattice values of Eqs.~(\ref{HLTRHS})
and~(\ref{HLTRHShat}),
we see that the lattice errors are factors of about 1.9, 1.2, 0.7 and 0.7
times the differences shown in Eq.~(\ref{BaBarKLOEdiff}) for $W'_{1,5}$,
$W'_{2,5}$, $\widehat{W}_{1,5}$ and $\widehat{W}_{2,5}$, respectively.
This implies that in order to use the EWSRs with the primed weights, an
improvement of a factor about 2 to 4 in the lattice errors would be needed,
while for the hatted weights, a very modest improvement would be sufficient
for the EWSRs to weigh in on the BaBar--KLOE discrepancy. A combination of
more precise lattice data and possibly a more fine-tuned design of the
sum-rule weight functions would help accomplishing this. We note that,
for this goal, the weight $W'_{2,5}$ was better ``designed'' than the
weight $W'_{1,5}$, and the weight $\widehat{W}_{2,5}$ similarly
better ``designed'' than the weight $\widehat{W}_{1,5}$. We also note
that, in both cases, the hatted weights outperform their primed
counterparts.

\begin{boldmath}
\section{\label{tau} Using hadronic $\t$-decay data for $a_\m^{\rm HVP}$}
\end{boldmath}

We now turn to a different application of the sum rules introduced in
Sec.~\ref{sumrules}: the use of spectral data obtained from non-strange
vector-current-induced hadronic $\t$-decay data in the determination
of $a_\m^{\rm HVP}$.

The idea of using $\tau$ data is straightforward in the isospin
limit. If isospin were an exact symmetry, one could replace
contributions to the $R$-ratio from $I=1$ exclusive modes in
$e^+e^-\to$~hadrons by the corresponding contributions implied by {the CVC (conserved vector
current) relation} and the related vector-channel ($G$-parity positive) exclusive-mode
hadronic $\tau$-decay distributions, for $s\leq m_\tau^2$. Doing so
is potentially useful for three reasons: (a) if, for a certain energy
region, the $\t$ data are more precise, this would help in improving
the precision of the dispersive determination of $a_\m^{\rm HVP}$;
(b) the $\tau$ data might shed light on the long-standing discrepancy
between BaBar and KLOE results for the $e^+e^-\to\p^+\p^-$ cross
sections observed in the $\r$ region discussed in Sec.~\ref{BaBarKLOE};
and (c) likewise, the $\tau$ data might shed light on the increasing
evidence for discrepancies between $R$-ratio and lattice based evaluations
of $a_\m^{\rm HVP}$ or closely related window quantities.

One of the reasons the current best dispersive determination of
$a_\m^{\rm HVP}$ does not employ hadronic $\t$-decay data is that, in the
real world, isospin is broken by electromagnetic corrections and the
difference in up and down quark masses. Typically, both isospin-breaking (IB)
effects are expected to be of order a percent or so (except possibly in the
region of narrow interfering resonances, where the effects may be larger).
Given the precision with which $a_\m^{\rm HVP}$ is obtained in the
dispersive approach, IB effects thus need to be brought under quantitative
control. Models for IB have been employed in the past \cite{review}, but
these are not under full phenomenological control, as evidenced by the fact
that they fail to account for the observed differences between the
experimental $\tau$ and electroproduction $\pi\pi$ distributions (see,
for example, Figs. 20 and 22 of Ref.~\cite{review}). Because of the
steady improvement in $R$-ratio data, and the problem of achieving an
understanding of the IB corrections one needs to apply to the $\tau$
$\pi\pi$ data, recent high-precision dispersive estimates of
$a_\m^{\rm HVP}$ do not involve $\t$-decay data at all.

There are, nevertheless, several reasons to revisit this topic. First,
lattice QCD may be able to produce relevant, first-principle, estimates
for IB effects that are useful in the comparison between $e^+e^-$-based
and $\t$-based data, as will be discussed below. Second, a new combination
of the sum of the $\p^-\p^0$, $2\p^-\p^+\p^0$ and $\p^-3\p^0$
contributions to the isovector, vector spectral function from ALEPH and
OPAL data has recently been obtained, producing an at-present-best
publicly available result for the sum of the contributions from these
modes \cite{AO}.\footnote{The reason only the sum is available is that
the ALEPH and OPAL covariance matrices for the two $4\p$ distributions
are highly singular, which can lead to biased results in the
data-combination procedure. The sum of $2\p$ and $4\p$ contributions
has a much better behaved covariance matrix and avoids this issue.}
Significant further improvements to the precision for this sum may,
moreover, be achievable using Belle II data.\footnote{An additional
practical issue is the need to ensure the radiative corrections applied to
$\tau$ data match those employed in the analysis of electroproduction data.
For more on this issue, and work in progress on evaluating the relevant
corrections, see Ref.~\cite{BrunoMeyer}.}

The basic idea is, using the weights $W_{m,n}$, $W_{m,n}^\prime$ or
$\widehat{W}_{m,n}$, to compare the weighted spectral integrals
computed with spectral data obtained from electroproduction and from
hadronic $\t$ decays. For any (sum of) $I=1$ channels, the SM difference
is caused by IB, and can be expressed as the corresponding difference of
the RHS of Eq.~(\ref{sumruleCt}) (or~(\ref{sumruleexpon})). In principle, the
RHS differences can be computed on the lattice, as we will discuss in
Sec.~\ref{IB}. The other important ingredient is the precision with which
the $W_{m,n}$-, $W_{m,n}^\prime$-, or $\widehat{W}_{m,n}$-weighted
spectral integrals can be evaluated. As we will see, choosing weights
like $W_{1,5}$ and $W_{2,5}$, which zoom in on the region of the
BaBar--KLOE discrepancy, the weighted spectral integrals of the
electroproduction- and $\t$-based data are already sufficiently
precise that results using current $\tau$ data will allow one to
weigh in on the BaBar--KLOE discrepancy, provided the required IB
corrections can be evaluated with comparable precision on the lattice.
We will discuss this point more quantitatively in Sec.~\ref{comptaue+e-}.

\subsection{\label{IB} Isospin breaking}

The main problem in comparing spectral data from electroproduction and
hadronic $\t$ decays using lattice data to compute IB effects in the
hadronic vacuum polarization $\P (Q^2)$ is that the lattice can only access
$\P(Q^2)$ for Euclidean $Q^2$.\footnote{Or, equivalently, $C(t)$ for
Euclidean $t$.} This is not a problem if one wants to compare fully
inclusive results for $a_\m^{\rm HVP}$, since representations exist
for $a_\m^{\rm HVP}$ both as a weighted integral of the $R$-ratio, $R(s)$,
over $s$\cite{Brodsky:1967sr,Lautrup:1968tdb,Gourdin:1969dm} and as a
weighted integral of $\hat{\P}(Q^2)$ over
$Q^2$\cite{Lautrup:1971jf,deRafael:1993za,Blum:2002ii}. It does, however,
become a problem if one wants to restrict oneself to contributions
from a subset of exclusive modes, and/or to restrict $s$ to a given region
$s_{\rm min}\le s\le s_{\rm max}$. Because of the kinematic restriction
$s\le m_\t^2$ for $\t$ decays, $s_{\rm max}$ by necessity has to be chosen
$\le m_\t^2$, whereas one may want to choose $s_{\rm min}$ larger than
threshold to reduce the relative statistical errors. Similar observations
apply to the sum rules of Sec.~\ref{sumrules}, in which, as for $a_\m^{\rm HVP}$,
the spectral integrals are evaluated from $s_{\rm th}$ to $\infty$.

The bulk of the dispersive contributions to $a_\m^{\rm HVP}$ come from
electroproduction data below $\sqrt{s}=1.937$~GeV, with $e^+e^-\to 2\p$,
$3\p$, $4\p$ and $K\bar{K}$ accounting for 72.8\%, 6.7\%, 4.9\% and
5.2\%, respectively, for a total of 89.7\%, according to
Ref.~\cite{KNT19}.

Contributions from all other exclusive modes below $\sqrt{s}=1.937$~GeV
represent a further 2.6\%, those from narrow charm and bottom resonances a
further 1.1\%, those from inclusive $R(s)$ data between $\sqrt{s}=1.937$
and $11.199$~GeV a further 6.3\%, and those (evaluated using pQCD) from
$\sqrt{s}$ above $11.199$~GeV a further 0.3\%.
The contributions from the $2\pi$, $3\pi$, $4\pi$ and $K\bar{K}$ exclusive
modes below $\sqrt{s}=1.937$~GeV, similarly, represent 98\% and 89\%,
respectively, of the full spectral integrals $I_{W_{1,5}}$ and $I_{W_{2,5}}$.

With contributions from the remaining exclusive modes below
$\sqrt{s}=1.937$~GeV lying at $s$ far removed from narrow interfering
resonances which might enhance IB, it seems reasonable to expect IB in
these residual exclusive-mode contributions to be very small. We
estimate these as of order 1\% of the 2.6\% total contribution from
these modes.

In the region above $\sqrt{s}=1.937$~GeV, away from the narrow charm and
bottom resonances, the experimental $R(s)$ data used in obtaining the 6.3\%
inclusive-region contribution of Ref.~\cite{KNT19} agrees with pQCD expectations
to significantly better than 10\%. With IB corrections well below 1\% in the
OPE representation of $\rho_{\rm EM}(s)$, estimating possible IB contributions
from the region above $\sqrt{s}=1.937$~GeV as less than 1\% of the 7.7\%
total from this region should represent a very conservative assessment.

With the 1\% or less estimates discussed above for IB in both the
contributions from exclusive modes other than $2\pi$, $3\pi$, $4\pi$ and
$K\bar{K}$ below $\sqrt{s}=1.937$~GeV and all contributions in the region
above $\sqrt{s}=1.937$~GeV, we conclude that IB contributions from these
sources should represent less than 0.1\% of the total for
$a_\m^{\rm HVP}$; a similar conclusion holds for  $I_{W_{1,5}}$ and
$I_{W_{2,5}}$. In summary, since the contributions from channels
other than $2\pi$, $3\pi$, $4\pi$ and $K\bar{K}$ below $\sqrt{s}=1.937$~GeV
and all contributions in the region above $\sqrt{s}=1.937$~GeV
represent less than $11\%$ of the totals for all of $a_\m^{\rm HVP}$,
$I_{W_{1,5}}$ and $I_{W_{2,5}}$, and it is safe to estimate IB for these
quantities as at most of order $1\%$, IB contributions should represent
of order $0.1\%$ or less of the total for all three quantities.
The uncertainty resulting from neglecting such contributions
can thus be safely neglected given the present state of the art.

In Ref.~\cite{disc} it was shown that the IB effect in the $I=1$ vector
$K\bar{K}$ channel is expected to be much smaller than the typical few
percent of the full $I=0+1$ sum, because $\t$-decay data for the $I=1$
contribution allows one to see that the $K\bar{K}$ electroproduction
cross sections are strongly dominated by the $I=0$ contribution from the
$\f$-peak region \cite{BaBarKKbar}. The IB effects on the small $I=1$
part of the $K\bar{K}$ channel should thus also be negligibly small.

It follows from the analysis of Ref.~\cite{disc} that, to first order in IB, IB
effects for all non-strange vector-channel $\tau$ exclusive-mode
contributions except those from the 2-pion and 4-pion modes are so small
that for the purpose of estimating IB corrections (apart from possible IB
effects in the $e^+e^-\rightarrow 3\pi$ distribution to be discussed in more
detail below) a fully inclusive lattice result for
\begin{equation}
\label{IBvacpol}
\P_{\rm IB}(Q^2)\equiv \frac{2}{\sqrt{3}}\,
\P_{38}(Q^2)+\left[\P_{33}(Q^2)-\half\P_{ud;V}(Q^2)\right]
\end{equation}
can be used to correct the sum of $2\pi$ and $4\pi$ contributions
obtained from non-strange vector-channel hadronic $\t$-decay data,
producing an alternate, combined $\t$-plus-lattice-IB determination for
the sum of $\p^+\p^-$, $2\p^+2\p^-$ and $\p^+\p^-2\p^0$ electroproduction
contributions.\footnote{A preliminary lattice study of some of the
contributions to Eq.~(\ref{IBvacpol}) already exists \cite{BILM,BrunoMeyer}.}

From the above discussion, it follows, using Eq.~(\ref{sumrulemom}),
again up to possible IB contributions
in the $e^+e^-\rightarrow 3\pi$ distribution which, to first order in
IB, would produce unwanted 3-pion contributions to the mixed-isospin,
$ab=38$ part of $\rho_{\rm EM}(s)$, and hence also to $\P_{38}(Q^2)$ in
Eq.~(\ref{IBvacpol}), that
\begin{eqnarray}
\label{sumruletau}
\D_{m,n}&\equiv&
\int_{s_{\rm th}}^{s_{\rm KNT}} ds\, W_{m,n}(s;\{Q_\ell^2\})
\r^{e^+e^-}_{2\pi +4\pi}(s)-
\int_{s_{\rm th}}^{s_{\rm KNT}} ds\, W_{m,n}(s;\{Q_\ell^2\})
\r^\t_{2\pi +4\pi}(s))\nonumber\\
&=&(-1)^m m_\t^{2(n-m-1)}\sum_{k=1}^n\frac{(Q_k^2+s_{\rm th})^m}
{\prod_{\ell\ne k}(Q_\ell^2-Q_k^2)}\,\P_{\rm IB}(Q_k^2)\ ,
\end{eqnarray}
where $s_{\rm KNT}=(1.937\ {\rm GeV})^2$ is the upper edge of
the exclusive-mode data region of Ref.~\cite{KNT19},
$\r^{e^+e^-}_{2\pi +4\pi}$ is the sum of the $2\pi$ and
$4\pi$ contributions to the electromagnetic current spectral function
implied by $2\pi$ and $4\pi$ electroproduction cross sections and
$\r^\t_{2\pi +4\pi}$ is the sum of the $2\p$ and $4\p$ contributions
to the consistently normalized vector $I=1$ spectral function implied by
non-strange vector-current-induced hadronic $\t$ decay distribution
results. Alternate versions of this sum rule in terms of $C_{\rm IB}(t)$
(\seef\ Eqs.~(\ref{sumruleCt}) or~(\ref{sumruleexpon})) can of course also
be employed. We will discuss the feasibility of this through concrete
examples in Sec.~\ref{comptaue+e-}, where we will evaluate the LHS
of Eq.~(\ref{sumruletau}) for the weights $W_{1,5}$ and $W_{2,5}$ using
data from Refs.~\cite{AO,KNT19}.

We now return to the issue of possible non-trivial IB effects in
the $e^+ e^-\rightarrow 3\p$ distribution, which complicates the
process of identifying the exclusive-mode contributions in the
dispersive integral over the electroproduction data to be replaced
by lattice-IB-corrected $\tau$ results. This complication arises because
there may be 3-pion exclusive-mode contributions to the spectral
function of the mixed-isospin $\Pi_{38}$ polarization. In the isospin
limit, the $3\p$ mode is pure $I=0$, and thus not affected by the
replacement of part of the $I=1$ component of the electromagnetic
spectral function by the corresponding part of the vector-channel,
$I=1$ $\t$ spectral function. With IB, however, the $38$ part of
Eq.~(\ref{IBvacpol}) will contain a 3-pion contribution, with
$\rho$-$\omega$ mixing, for example, inducing a contribution to
the $38$ spectral function via $e^+e^-\to\r\to\omega\to 3\p$.
Applying the inclusive lattice IB correction to the weighted
$\t$ $2\p+4\p$ integral and using the result to replace the
corresponding $2\p+4\p$ contribution to the weighted electroproduction
integral, the modified all-exclusive-mode sum would then include this
$3\p$ IB component twice, once in the inclusive lattice IB correction,
and once in the $3\p$ contribution produced by use of the experimental
$e^+e^-\rightarrow 3\p$ cross sections. The result of Ref.~\cite{KNT19}
for the contribution of the $3\p$ channel to $a_\m^{\rm HVP}$,
$46.73(94)\times 10^{-10}$, is sufficiently large that, were the
associated IB $3\p$ contribution to be larger than naively expected,
say at a level a few to several percent of the full $3\p$ contribution,
a controlled estimate of its size might be required to make an inclusive
lattice $\P_{\rm IB}(Q^2)$ determination, however precise, of actual
numerical use in performing an IB correction of $\t$-decay data.

Fortunately, experimental information is now available on what should
be the dominant contribution to this $3\p$ double counting, with
recent BaBar results~\cite{BABAR:2021cde} providing $>6\sigma$
evidence for an IB $e^+ e^-\rightarrow \rho \rightarrow 3\pi$
contribution to the $e^+ e^-\rightarrow 3\pi$ amplitude. The
interference of this contribution with the isospin-conserving (IC)
$e^+ e^-\rightarrow\omega\rightarrow 3\pi$,
$e^+ e^-\rightarrow\phi\rightarrow 3\pi$,
$e^+ e^-\rightarrow \omega^\prime\rightarrow 3\pi$ and
$e^+ e^-\rightarrow \omega^{\prime\prime}\rightarrow 3\pi$ contributions
produces IB contributions to the cross section, and hence
IB contributions to the weighted integrals over the $3\pi$ distribution
involving the weights discussed in this paper. The normalization of these
interference contributions is fixed by the square root of the fitted
$\rho\rightarrow 3\pi$ branching fraction, $B(\rho\rightarrow 3\pi)$, to
which these contributions are proportional. Taking the integrals up to
the upper edge, $s=(1.937\ {\rm GeV})^2$, of the KNT19 exclusive-mode
region to be specific, the IB contributions implied by the preferred
BaBar cross-section fit, as detailed in App.~\ref{IB3pion}, turn out to
represent $-1.2(1.2)\%$, $-1.1(1.1)\%$ and $-1.6(1.6)\%$ of the
corresponding full $3\pi$ contributions to $a_\mu^{\rm HVP}$,
$I_{W_{1,5}}$ and $I_{W_{2,5}}$, respectively.

In the $a_\m^{\rm HVP}$ case, the resulting $3\pi$-double-counting
correction, $+0.54(0.54)\times 10^{-10}$ (from Eq.~(\ref{ib3piintestimates})),
represents $+0.08(0.08)\%$ of the $a_\m^{\rm HVP}$ total. At present,
a $\pm 0.54\times 10^{-10}$ uncertainty on the $\tau$-modified
alternate determination of $a_\m^{\rm HVP}$ is sufficiently small that
replacing electroproduction data by $\t$-based data for the $2\p$ and
$4\p$ channels is potentially useful for the purposes of
investigating, for example, the BaBar--KLOE discrepancy, where the
discrepancy between BaBar and KLOE $2\pi$ contributions from the region
$0.6\ {\rm GeV}< \sqrt{s}<0.9$~GeV~\cite{KNT19}, $(9.8\pm 3.5)\times
10^{-10}$, is much larger than the $3\pi$ double-counting-correction
uncertainty. In Sec.~\ref{comptaue+e-} we will show that a similar
conclusion holds for the $W_{1,5}$ and $W_{2,5}$ weight cases:
the $\pm 0.00041$ and $\pm 0.00012$ uncertainties on the $W_{1,5}$ and
$W_{2,5}$ $3\p$-double-counting corrections obtained in App.~\ref{IB3pion}
are once more much smaller than the central values and experimental errors
on the corresponding electroproduction-$\tau$ spectral integral differences.

A more detailed discussion of the IB contributions to the weighted
$3\pi$ integrals may be found in Appendix~\ref{IB3pion}.

\begin{boldmath}
\subsection{\label{comptaue+e-} Comparison of $\t$-based and $e^+e^-$-based spectral
integrals}
\end{boldmath}
We now turn to the evaluation of the $2\p +4\p$ contributions to
the spectral integrals appearing in Eq.~(\ref{sumruletau}), for the examples
$W_{m,n}=W_{1,5}$ and $W_{2,5}$. The $2\pi +4\pi$ data of Ref.~\cite{KNT19}
is used in the first integral and that of Ref.~\cite{AO} in the second integral.
We also consider the $3\p$ channel, in view of the discussion at the end
of Sec.~\ref{IB}. The discussion is restricted to contributions from the
exclusive-mode region of Ref.~\cite{KNT19}, $s\le s_{\rm KNT}=(1.937\ {\rm GeV})^2$,
since $2\p$, $3\p$ and $4\p$ contributions above that point form part of the
multi-mode inclusive contribution, dealt with already in Sec.~\ref{IB}.
Since no lattice data is currently available to evaluate the
RHS of either Eq.~(\ref{sumruletau}) or its EWSR analogue, we restrict
our attention to the $W_{1,5}$ and $W_{2,5}$ RWSR examples, and do
not consider the EWSR analogues. Once suitable IB lattice data become
available, the analogous EWSR cases may become of interest, for
the reasons explained in Sec.~\ref{exsumrule2}.

Our goal here is to see whether the two integrals on the LHS of
Eq.~(\ref{sumruletau}) can be evaluated with sufficient precision to
make this sum rule of potential use. Of relevance to investigating
this question is the size of the uncertainty on the spectral integral
differences appearing on the LHS of Eq.~(\ref{sumruletau}) relative to (i) the
uncertainties on estimates for the corresponding $3\pi$-double-counting
corrections, and (ii) variations in the spectral integral differences
themselves, induced, for example, by the BaBar--KLOE $\pi\pi$ discrepancy.

We start with the electroproduction-based $W_{1,5}$- and $W_{2,5}$-weighted
integrals over the $2\p$, $3\p$ and $4\p$ contributions to $\rho_{\rm EM}(s)$,
results for which are shown in Table~\ref{EMW1525}. We use
$s_{\rm th}=4m_\p^2$.

\begin{table}[t]
\begin{center}
\begin{tabular}{|l||l|l| c || l | l | c |}
\hline
channel & $I_{W_{1,5}}(s_{\rm KNT})$ & $I_{W_{1,5}}(s_\t)$
& $\frac{I_{W_{1,5}}(s_\t)}{I_{W_{1,5}}(s_{\rm KNT})}$
& $I_{W_{2,5}}(s_{\rm KNT})$  & $I_{W_{2,5}}(s_\t)$
& $\frac{I_{W_{2,5}}(s_\t)}{I_{W_{2,5}}(s_{\rm KNT})}$\\
\hline
$\p^+\p^-$ & 0.3864(14) & 0.3863(14) & 1.00 & 0.05713(19)
& 0.05710(19) &1.00 \\
$2\p^+2\p^-$& 0.005743(81) & 0.005450(77) & 0.95 & 0.003568(49)
& 0.003267(45) & 0.92 \\
$\p^+\p^-2\p^0$ & 0.00813(33) & 0.00772(32) & 0.95 & 0.00464(19)
& 0.00422(17) & 0.91 \\
\hline
$2\p+4\p$ & 0.4002(14) & 0.3995(14) & 1.00 & 0.06534(27)
& 0.06459(26) & 0.99 \\
\hline
$\p^+\p^-\p^0$ & 0.03880(81) & 0.03877(81) & 1.00 & 0.00798(15)
& 0.00794(15) & 1.00 \\
\hline
\end{tabular}
\end{center}
\vspace*{-3ex}
\begin{quotation}
\floatcaption{EMW1525}{{\it Electroproduction-based exclusive-mode
spectral integrals with weights $W_{1,5}$ (left of the double
vertical line) and $W_{2,5}$ (right of the double vertical line).
The 2nd and 5th columns show the integrals computed up to
$s=s_{KNT}=(1.937\ {\rm GeV})^2$, the 3rd and 6th columns the same
integrals computed up to $s=s_\t=3.0574$~{\rm GeV}$^2$.
}}
\end{quotation}
\vspace*{-4.5ex}
\end{table}

The data files containing the exclusive-mode contributions to $R(s)$ provided
by the authors of Ref.~\cite{KNT19} extend up to $s=s_{\rm KNT}=3.7520$~GeV$^2$.
Results for the exclusive-mode, $s_{\rm KNT}$-truncated $X=\pi^+\pi^-$,
$2\pi^+2\pi^-$, $\pi^+\pi^-2\pi^0$ and $\pi^+\pi^-\pi^0$, $W_{1,5}$- and
$W_{2,5}$-weighted spectral integrals obtained using this input, are denoted
$I^X_{W_{1,5}}(s_{\rm KNT})$ and $I^X_{W_{2,5}}(s_{\rm KNT})$ in what follows,
and listed in Columns 2 and 5 of Table~\ref{EMW1525}.

The upper endpoints of the $\t$-based spectral integrals are, in contrast,
limited by the largest $s$ for which the $\tau$-based spectral function of
Ref.~\cite{AO} is available, which is $s\equiv s_\t=3.0574$~GeV$^2$, slightly
below $m_\tau^2$. To compare to the resulting $s_\tau$-truncated $\t$-based
integrals, we thus also require values for $s_\t$-truncated versions of
the exclusive-mode electroproduction-based integrals. These are denoted
$I^X_{W_{1,5}}(s_\tau )$ and $I^X_{W_{2,5}}(s_\tau )$ and listed in columns
3 and 6 of Table~\ref{EMW1525}. As can be seen from the table, the
$s_\tau$-truncated integrals constitute more than $99\%$ of the corresponding
$s_{\rm KNT}$-truncated versions for the $2\p$ and $3\p$ modes and more than
90\% of the $s_{\rm KNT}$-truncated versions for the two $4\p$ modes.

The $s_\tau$-truncated $2\p+4\p$ $\tau$-based integrals to be compared to
the $R$-ratio-based analogues of Table~\ref{EMW1525} are obtained
using the $2\p+4\p$ $\t$-based spectral function of Ref.~\cite{AO}.
The results,
\begin{eqnarray}
\label{tauW1525tau}
I_{W_{1,5}}^{\t,2\p+4\p}(s_\t)&=&0.4103(22)\ ,\\
I_{W_{2,5}}^{\t,2\p+4\p}(s_\t)&=&0.06693(22) \ ,\nonumber
\end{eqnarray}
differ from the electroproduction results of Table~\ref{EMW1525},
\begin{eqnarray}
\label{tauW1525R}
I_{W_{1,5}}^{2\p+4\p}(s_\t)&=&0.3995(14)\ ,\\
I_{W_{2,5}}^{2\p+4\p}(s_\t)&=&0.06459(26) \ ,\nonumber
\end{eqnarray}
by $4.1\s$ and $6.7\s$, respectively, for the $W_{1,5}$ and $W_{2,5}$ cases.
The $s_\tau$-truncated contributions to the differences appearing
on the LHS of the $W_{1,5}$ and $W_{2,5}$ versions of Eq.~(\ref{sumruletau}),
\begin{eqnarray}
\label{tauW1525Rdiffs}
I_{W_{1,5}}^{2\p+4\p}(s_\t)-I_{W_{1,5}}^{\t,2\p+4\p}(s_\t)
&=&\, -0.0108(26)\ ,\\
I_{W_{2,5}}^{2\p+4\p}(s_\t)-I_{W_{2,5}}^{\t,2\p+4\p}(s_\t)
&=&\, -0.00233(35) \ ,\nonumber
\end{eqnarray}
are thus determined with $\sim 24\%$ and $\sim 15\%$ precision.
We note that the errors in Eq.~(\ref{tauW1525Rdiffs}) are significantly
smaller than the central values of the BaBar--KLOE discrepancies in
Eq.~(\ref{BaBarKLOEdiff}).
The analogous values for $a_\m^{\rm HVP}$ itself are
\begin{eqnarray}
\label{amudiff}
a_\m^{2\p+4\p}(s_\t)&=&535.0(2.0)\times 10^{-10}\ ,\\
a_\m^{\t,2\p+4\p}(s_\t)&=&552.4(5.0)\times 10^{-10}\ ,\nonumber\\
a_\m^{2\p+4\p}(s_\t)-a_\m^{\t,2\p+4\p}(s_\t)&=&-17.4(5.4)\times 10^{-10}
\ .\nonumber
\end{eqnarray}
We comment on the $\t$-based value in Eq.~(\ref{amudiff}) (\ie, the second line
of this equation).   The $\t$-based data are very sparse near the 2-pion threshold end of
the spectral function, and a comparison with the $R$-ratio-based data (which are much denser in the threshold region) suggests that the trapezoidal interpolation of the $\t$-based data may
overestimate the near-threshold, $\t$-based contribution to $a_\m^{\t,2\p+4\p}(s_\t)$.
A rough estimate of this effect can be obtained by replacing the contribution
to $a_\m^{\t,2\p+4\p}(s_\t)$
between threshold and $s_{\rm ChPT}\equiv s=(0.305$~GeV)$^2$ by the ChPT-based 2-pion
contribution from Ref.~\cite{KNT19}, and then employing trapezoidal integration
above $s_{\rm ChPT}$.  Doing so would lead to a downward shift of
$0.90(42)\times 10^{-10}$ of the value $552.4(5.0)\times 10^{-10}$ in
Eq.~(\ref{amudiff}).  This small additional near-threshold uncertainty in the $\t$ result for
$a_\m^{2\p +4\p}$ is a consequence of the enhancement of low-$s$
contributions by the $a_\m^{\rm HVP}$ kernel. The weights $W_{1,5}$ and
$W_{2,5}$, in contrast, produce no such low-$s$ enhancement (and
in fact strongly suppress near-threshold contributions). The uncertainty
in the low-$s$ $\t$-based $2\p$ spectral distribution resulting from the coarseness
of the $\tau$ data near threshold thus has negligible impact on the results
in Eq.~(\ref{tauW1525tau}).
Since weighted spectral integrals
with the weights $W_{1,5}$ and $W_{2,5}$ are the main focus of this paper,
we do not pursue this issue further here.

Of course, before we can meaningfully compare the $\tau$- and
electroproduction-based integrals, we need to take the RHS, \ie, the
projected lattice-based IB correction, into account.   Since this
contribution is inclusive, we have to deal with the fact that the
$s_\tau$-truncated spectral integrals whose differences we are actually
able to determine from data are smaller than the complete spectral
integrals. While we have already argued in Sec.~\ref{IB} that IB contributions
from modes other than $2\pi$ and $4\pi$ below $s=s_{\rm KNT}$, and
from all modes above $s=s_{\rm KNT}$, can be safely neglected, we still
need to address the size of possible IB contributions from the
$2\pi$ and $4\pi$ modes in the region $s_\tau <s< s_{\rm KNT}$.
For the $2\p$ modes, this is not an issue, as the integrals up to $s_\t$
capture essentially the full contribution up to $s_{\rm KNT}$, for both
weights we consider here. The $s_\tau$-truncated $4\p$ integrals constitute
$95\%$, respectively, $91$-$92\%$ of the full $4\pi$ integrals, for the
weights $W_{1,5}$ and $W_{2,5}$. In this high-$s$, $s>s_\t$ region, we
expect IB to be of order $1\%$ of the corresponding $4\pi$ totals. With
this estimate, the IB contribution missed as a result of truncating the
$W_{1,5}$ and $W_{2,5}$ integrals at $s=s_\t$ rather than $s_{\rm KNT}$
are expected to be (i) in the $W_{1,5}$ case, of order $1\%$ of $5\%$, or
$0.05\%$, of the corresponding full $s_{\rm KNT}$-truncated $4\pi$
contribution, and hence of order $0.000007$, and (ii) in the
$W_{2,5}$ case, of order $1\%$ of $9\%$, or $0.09\%$, of the corresponding
full $s_{\rm KNT}$-truncated $4\pi$ contribution, and hence of order
$0.000007$ as well. These estimated higher-$s$ $2\pi +4\pi$ IB
contribution effects, missed when one considers the differences of
electroproduction- and $\tau$-based $2\pi +4\pi$ integrals only up
to $s=s_\tau$, are thus more than an order of magnitude smaller than
the uncertainties, $0.00041$ and $0.00012$, on the corresponding
$3\pi$-double-counting corrections estimated in App.~\ref{IB3pion}, which
are, themselves, significantly smaller than both the central values
and experimental errors of the Eqs.~(\ref{tauW1525Rdiffs}) results,
$-0.0108\pm 0.0026$ and $-0.00233\pm 0.00035$, for the $W_{1,5}$-
and $W_{2,5}$-weighted, $s=s_\tau$-truncated experimental
electroproduction-$\t$ $2\pi +4\pi$ spectral integral differences,
indicating that a sufficiently precise lattice determination of the
corresponding RHSs of Eq.~(\ref{sumruletau}) will, indeed, make these sum rules
useful for investigating current electroproduction- and $\t$-based integral
results. It remains of course crucial that a lattice estimate of the
inclusive IB correction reach a precision commensurate with the
precision with which the spectral integral differences in
Eqs.~(\ref{tauW1525Rdiffs}) have been obtained.

\vspace{1cm}
\section{\label{conclusions} Conclusions}
At present, there are several discrepancies in the computation of
$a_\m^{\rm HVP}$ that limit our ability to compare a SM-based estimate
for $a_\m$ with the experimentally measured value. Most recently, a
puzzling discrepancy has emerged between data-driven and lattice
evaluations of the RBC/UKQCD intermediate window quantity,
suggesting that a lattice-based value for $a_\m^{\rm HVP}$ may bring
the SM value for $a_\m$
much closer to the experimental value. This may confirm the lattice
result for $a_\m^{\rm HVP}$ found by the BMW collaboration, which is
$2.1\s$ higher than the data-driven value. The discrepancy for the
intermediate window turns out to be about half the total difference
between the experimental and SM values for $a_\m$, when for the latter
the data-driven white-paper value for $a_\m^{\rm HVP}$ of Ref.~\cite{review}
is used.

Another discrepancy results from the long-standing difference between
the BaBar and KLOE measured spectral distributions in the two-pion channel,
in the region around the $\r$ mass. Taking this discrepancy at face value,
\ie, considering the difference between the values obtained using either the
BaBar or the KLOE data, leads to a difference of about $10^{-9}$ in
$a_\m^{\rm HVP}$, which is more than half the difference between the
experimental and white-paper values for $a_\m$.

Given this puzzling state of affairs, it is important to develop methods that
allow for detailed comparisons, zooming in on specific regions in $s$.
The sum rules developed in this paper provide a tool for such investigations.
The RWSRs of Sec.~\ref{sumrule1} are based on weighted spectral integrals with
an adjustable narrow-weight function defined directly as a function of $s$.
The quantities defined by such sum rules complement the window
quantities of Ref.~\cite{RBC}, which are defined as a function of Euclidean
time, and translate into rather wide windows as a function of
$s$.\footnote{Our method differs from that of Ref.~\cite{Rwindow}, which
attempts to narrow the window as a function of $s$ by taking linear
combinations of windows defined as a function of Euclidean time.}

In Sec.~\ref{sumrule2} we modified the RWSR approach by borrowing ideas
from Ref.~\cite{HLT}, proposing a different set of sum rules, with weight
functions that are linear combinations of simple exponential functions
of $\sqrt{s}$. This class of sum rules has the advantage that the lattice
side involves the lattice correlator for the vacuum polarization at a
set of Euclidean $t$ values which are chosen by hand and which hence
allow one to avoid contributions from the high-$t$ region, where lattice
errors are large. This aids in the optimization of the precision of the
lattice side of the sum rule.  A key observation is that the
weights used for the RWSRs can, moreover, be used as molds to cast the
exponential weights, thus retaining the advantage of RWSRs. We note that our
goal is not to reconstruct spectral data from the lattice, but rather
to compare appropriately weighted, moderately localized versions
of existing experimental spectral data with correspondingly weighted
lattice data. Our goal is thus simpler than the goal of Ref.~\cite{HLT}.
In particular, the casts are not used in any approximations; once a
useful cast has been designed, it can be used in an exact EWSR, and
the underlying mold discarded. In fact, there is considerable
flexibility in designing EWSRs. An example of this flexibility is
provided by the hatted weight functions based on Eq.~(\ref{Ahat}).
Two key points are worth reiterating here. First, weights $W(s)$ very
similar as functions of $s$ may produce significantly different relative
errors on the lattice sides of the associated EWSRs. And second, given
that the sum rules corresponding to different weights are all exact, one
is free to choose, from any set of such similar weights, the one that
produces the most stringent dispersive-lattice comparison. In the examples
explored above, this would be the hatted EWSR weights.
Additional ideas for designing practical exponential weights that
are narrow as a function of $s$, moreover, almost certainly remain
to be explored.

At first glance, it is not obvious that these new methods will be
practically useful, as there are reasons to worry that the lattice sides
of our sum rules will typically have large errors. To investigate this
worry, we studied, in
Sec.~\ref{comparison}, two rational weights,
$W_{1,5}$ and $W_{2,5}$, and their exponential cousins, numerically,
using data from Ref.~\cite{KNT19} on the data side, and from Ref.~\cite{ABGP22}
on the lattice side. We found that especially the exponential weights
perform quite well for an investigation of, for example, the
BaBar--KLOE discrepancy. With the projected increase in the precision
of data from the lattice, and with more fine-tuning of the weights
of Sec.~\ref{sumrule2}, we believe that the new tools provided in this paper
are likely to prove quite useful for the investigation of
currently existing discrepancies.

In Sec.~\ref{tau} we applied these ideas in a somewhat different context,
the comparison between contributions to $a_\m^{\rm HVP}$ from $R$-ratio
data and hadronic $\t$-decay data. In the past, such comparisons were
difficult to carry out because of the lack of a reliable method for estimating
isospin-breaking effects. We derived a sum rule allowing for the comparison
between $R$-ratio and hadronic $\t$-decay based data in which the
required IB effects can, to a good approximation, be obtained from the
lattice. In particular, we showed that if the two- and four-pion channels are
included in the $R$-$\t$ comparison, IB effects in other channels are
small enough that IB effects can be reliably incorporated by an inclusive
lattice computation, thus avoiding the difficulties associated with
obtaining exclusive data from the lattice. Using the weights of
Sec.~\ref{comparison}, we demonstrated that this comparison can become
practical with sufficiently precise lattice data for the inclusive
IB correction.

We conclude by emphasizing that this paper is a method paper. The main
reason is that we do not have access to lattice data for the complete
hadronic vacuum polarization, or for the IB part of it needed in
Sec.~\ref{tau}. We have not optimized the concrete examples of
Sec.~\ref{comparison} and Sec.~\ref{tau} for the specific goals for which
they may be used. We leave the optimization of these new classes of
sum rules to future work, and here just comment that this optimization
will depend on the specific application.

\vspace{3ex}
\noindent {\bf Acknowledgments}\\

We thank Alex Keshavarzi for making all exclusive-mode data underlying
Ref.~\cite{KNT19} available to us. We thank Mattia Bruno and Max Hansen for
useful discussions, and Marcus V. Rodrigues for his participation
at an early stage of this work. DB's work was supported by the S\~ao Paulo
Research Foundation (FAPESP) Grant No. 2021/06756-6 and by CNPq Grant
No. 308979/2021-4. MG is supported by the U.S.\ Department of Energy, Office
of Science, Office of High Energy Physics, under Award DE-SC0013682. The work
of KM is supported by a grant from the Natural Sciences and Engineering
Council of Canada. SP is supported by the Spanish Ministry of Science,
Innovation and Universities (project
PID2020-112965GB-I00/AEI/10.13039/501100011033)
and by Grant 2017 SGR 1069. IFAE is partially funded by the CERCA program
of the Generalitat de Catalunya.

\appendix
\section{\label{precision} Relation between weights and precision}

Consider the weights of Eq.~(\ref{srweight}), without the extra factors of
$m_\tau$:
\begin{equation}
W_{m,n}(s)= {\frac{(s-s_{\rm th})^m}{\prod_{k=1}^n(s+Q_k^2)}}\ ,
\label{productform}
\end{equation}
where $0<Q_1^2<Q_2^2<\cdots <Q_n^2$. What we will discuss in this
appendix is the issue of the cancellations involved when this weight
is used in the sum-rule approach discussed in the main text.

The function $W_{m,n}(s)$ can be re-written in the alternate partial-fraction
representation
form
\begin{equation}
W_{m,n}(s)=\sum_{k=1}^n {\frac{c_k}{(s+Q_k^2)}}\ ,
\label{sumform}
\end{equation}
where the $c_k$ are easily seen to be
\begin{equation}
c_k={\frac{(-1)^m(Q_k^2+s_{\rm th})^m}{\prod_{\ell\ne k}(Q_\ell^2-Q_k^2)}}\ .
\label{cs}
\end{equation}
By  expanding in $1/s$ for large $s$, it is
straightforward to work out a number of relations satisfied by the
$c_k$, as follows. In the form Eq.~(\ref{productform}), it is clear that
\begin{equation}
W_{m,n}(s) = {\frac{1}{s^{n-m}}}\, +  \co\left( {\frac{1}{s^{n-m+1}}}\right)\ ,
\label{expand}
\end{equation}
where the expansion is convergent for $s>Q_N^2$. In contrast, expanding the
form Eq.~(\ref{sumform}), the large $s$ behavior is
\begin{equation}
W_{m,n}(s) = \sum_{k=1}^n c_k\, {\frac{1}{s}}\,
\sum_{\ell=0}^\infty (-1)^\ell \left({\frac{Q_k^2}{s}}\right)^\ell\ .
\label{expand2}
\end{equation}
Comparing the two large $s$ expansions, it follows that the
$c_k$ satisfy the relations
\begin{equation}
\sum_{k=1}^n c_k Q_k^{2\ell}=0\ ,\qquad \ell=0,\dots ,n-m-2\ .
\label{ckconstraints}
\end{equation}

The general dispersive sum rules involving these weights,
assuming $n$ is large enough that the weighted spectral integral
converges, are of course Eq.~(\ref{sumrulemom}), without the factors of
$m_\t$:
\begin{equation}
\label{prodpoleswtsr}
\int_{s_{\rm th}}^\infty ds\, W_{m,n}(s)\, \rho(s)=
\sum_{k=1}^n c_k\, \Pi (Q_k^2)
=(-1)^m\sum_{k=1}^n {\frac{(Q_k^2+s_{\rm th})^m}
{\prod_{\ell\ne k}(Q_\ell^2-Q_k^2)}}
\, \Pi (Q_k^2)\ .
\end{equation}

The constraints of Eq.~(\ref{ckconstraints}) allow us to understand the
cancellations involved in forming the sum that appears on the right-hand
side of Eq.~(\ref{prodpoleswtsr}). $\Pi (Q^2)$ is analytic on the $Q^2>0$ axis,
and hence has a convergent Taylor series expansion around any point on
that axis with radius of convergence the distance from that point to
the start of the cut at $Q^2\, =\, -s_{\rm th}$. For illustration in
what follows let us expand about the midpoint of the interval
containing all pole locations, $\tilde{Q}^2\equiv (Q_1^2+Q_n^2)/2$.
All of the $Q_k^2$ then lie in the region of convergence of the Taylor
expansion, and we have
\begin{equation}
\Pi (Q_k^2) = \sum_{\ell =0}^\infty {\frac{1}{\ell!}}\, {\frac
{d^\ell\P(Q^2)}{d(Q^2)^\ell}}\Bigg|_{Q^2=\tilde{Q}^2}
\left( Q_k^2-\tilde{Q}^2\right)^\ell\ .
\label{expandPi}
\end{equation}
The (first version of the) right-hand side of Eq.~(\ref{prodpoleswtsr}), then
becomes
\begin{equation}
\label{srtaylor}
\sum_{\ell =0}^\infty {\frac{1}{\ell!}}\, {\frac
{d^\ell\P(Q^2)}{d(Q^2)^\ell}}\Bigg|_{Q^2=\tilde{Q}^2}
\sum_{k=1}^n c_k \left(Q_k^2-\tilde{Q}^2\right)^\ell\ ,
\end{equation}
and the constraints of Eq.~(\ref{ckconstraints}) imply that terms in the last
factor vanish for $\ell = 0,\dots ,n-m-2$. The terms involving
the derivatives of order $0$ through $n-m-2$ of $\Pi (Q^2)$ with
respect to $Q^2$ at $Q^2=\tilde{Q}^2$ thus vanish, and the first
surviving terms are those involving the $(n-m-1)$-th derivative.

There is thus significant cancellation on the right-hand side of the
dispersive sum rule for weights with the product-of-pole rational structure
of Eq.~(\ref{productform}),
and this cancellation gets stronger with increasing $n-m$. This will
lead to errors on a lattice evaluation of the right-hand side which
will typically increase with increasing $n-m$. This growth of lattice
errors with increasing $n-m$ was seen already in the dispersive analysis
used to extract $|V_{us}|$ from the experimental strange $\t$-decay
distribution, which employed similar product-of-pole weights,
though with a constant numerator. There it was found that lattice
errors could be kept under good control for weights with 3, 4, or 5
poles, but the errors did grow as the number of pole factors
increased \cite{Kim}.

\section{\label{IB3pion} Isopin breaking in the 3-pion channel}
Experimental information on IB in the 3-pion channel is provided by BaBar's
vector-meson-dominance (VMD) model fit to its recent high-precision
$e^+ e^-\rightarrow 3\pi$ cross sections, detailed in Ref.~\cite{BABAR:2021cde}.
The VMD model for the amplitude is a sum of IC $\omega$, $\phi$,
$\omega^\prime$ and $\omega^{\prime\prime}$ resonance contributions,
supplemented by an IB $\rho$ contribution and provides an excellent fit
(with $\chi^2/{\rm dof} = 136/129$) to the experimental cross-sections in the
region $E_{\rm CM}\le 1.8$ GeV. The explicit model forms of the resonance
contributions are as specified in Refs.~\cite{BABAR:2021cde,Achasov:2003ir}.
The fit confirms the necessity of including the IB $\rho$ contribution
at the $>6\sigma$ level. The model includes what should be the dominant
(resonance-enhanced) IB contributions to the cross section in the region
from threshold to slightly above the $\phi$ resonance peak, namely the
effects of $\rho$-$\omega$ and $\rho$-$\phi$ interference. It also
includes sub-leading IB effects in the form of $\rho$-$\omega^\prime$ and
$\rho$-$\omega^{\prime\prime}$ interference contributions from the region
above the $\phi$ peak. With no IB contributions to the amplitude from
the excited $\rho$ resonances, however, it will miss IB contributions
in this higher-$s$ region from, {\eg}, $\rho^\prime$-$\omega^\prime$ and
$\rho^\prime$-$\omega^{\prime\prime}$ interference. Such contributions
will be suppressed by the fall-off with $s$ of the weights considered in
this paper and are thus also expected to be numerically sub-dominant.

In what follows, we take, as our estimates for the IB contributions
to weighted, $s\le s_{\rm KNT}$, $3\pi$ spectral integrals, the results
produced using the full BaBar VMD fit. As noted above, with the VMD model
omitting terms which would model, {\it e.g.}, IB $\rho^\prime$ and
$\rho^{\prime\prime}$ contributions to the $3\pi$ amplitude, such
estimates will miss some contributions from the region above the $\phi$ peak.
We will discuss IB in this region in more detail below and demonstrate
that such missing contributions are expected to be much smaller than the
uncertainties on the dominant low-$s$ $\rho$-$\omega$ plus $\rho$-$\phi$
interference contributions, and hence that the estimates for the IB
contributions to the variously weighted, $s\le s_{\rm KNT}$, $3\pi$
spectral integrals obtained using the BaBar VMD model fit are expected
to be reliable within their stated errors.

The BaBar paper~\cite{BABAR:2021cde} contains full results for the
central values and errors (though not the correlations) of the fit
parameters governing the $\omega$, $\phi$ and $\rho$ contributions to
the amplitude, but not those governing the $\omega^\prime$ and
$\omega^{\prime\prime}$ contributions. It is thus possible to determine
the $s$-dependence of the combined $\omega +\phi + \rho$ contribution
to the VMD model representation of the cross section, but not that of
the full VMD model. BaBar has, however, provided a table of central
values of the full model representation at the midpoints of the BaBar
experimental bins, both with the $\rho$ contribution included and with
that contribution turned off~\cite{thanksvdruzhinin}. This provides us
with the $s$-dependence of the central fit values of the IB part of the
VMD representation of the cross section and allows us to evaluate the
central values of the variously weighted integrated versions of
the IB $3\pi$ spectral distribution of interest in this paper.

In the discussions which follow, we will denote the $3\pi$ spectral
integrals up to $s=s_{\rm max}$ with the $a_\mu^{\rm HVP}$, $W_{1,5}$
and $W_{2,5}$ weights by $a_\mu^{{\rm HVP};\, 3\pi}\left( s_{\rm max}\right)$,
$I^{3\pi}_{W_{1,5}}\left( s_{\rm max}\right)$ and
$I^{3\pi}_{W_{2,5}}\left( s_{\rm max}\right)$.
The corresponding IB contributions obtained using BaBar's fitted VMD
model, again up to $s=s_{\rm max}$, are similarly denoted
$\left[ a_\mu^{{\rm HVP};\, 3\pi}\left( s_{\rm max}\right)
\right]_{\rm IB}^{\rm VMD}$, $\left[ I^{3\pi}_{W_{1,5}}\left(
s_{\rm max}\right) \right]_{\rm IB}^{\rm VMD}$ and
$\left[ I^{3\pi}_{W_{2,5}}\left( s_{\rm max}\right)
\right]_{\rm IB}^{\rm VMD}$.
Finally, the sums of the corresponding fully known IB VMD $\rho$-$\omega$
and $\rho$-$\phi$ interference contributions from the region up to just
above the $\phi$ peak (which we characterize, to be specific, as
$s\le s_\phi \equiv (m_\phi + 4\Gamma_\phi )^2$) are denoted by
$\left[ a_\mu^{{\rm HVP};\, 3\pi} \left( s_\phi \right)
\right]_{\rm IB}^{\rho\omega\phi ;\, {\rm VMD}}$,
$\left[ I^{3\pi}_{W_{1,5}}\left( s_\phi \right)
\right]_{\rm IB}^{\rho\omega\phi ;\, {\rm VMD}}$ and
$\left[ I^{3\pi}_{W_{2,5}}\left(s_\phi \right)\right]_{\rm IB}
^{\rho\omega\phi ;\, {\rm VMD}}$, respectively.

With this notation, we find the low-$s$ IB contributions,
$\left[ a_\mu^{{\rm HVP};\, 3\pi} \left(s_\phi \right)
\right]_{\rm IB}^{\rho\omega\phi ;\, {\rm VMD}}$,
$\left[ I^{3\pi}_{W_{1,5}}\left( s_\phi \right)
\right]_{\rm IB}^{\rho\omega\phi ;\, {\rm VMD}}$ and
$\left[ I^{3\pi}_{W_{2,5}}\left( s_\phi \right)\right]_{\rm IB}
^{\rho\omega\phi ;\, {\rm VMD}}$ to represent $80\%$, $87\%$ and $76\%$,
respectively, of the corresponding full IB estimates,
$\left[ a_\mu^{{\rm HVP};\, 3\pi}\left( s_{\rm KNT}\right)
\right]_{\rm IB}^{\rm VMD}$,
$\left[ I^{3\pi}_{W_{1,5}}\left( s_{\rm KNT}\right)
\right]_{\rm IB}^{\rm VMD}$ and
$\left[ I^{3\pi}_{W_{2,5}}\left( s_{\rm KNT}\right)
\right]_{\rm IB}^{\rm VMD}$, in line with the expectation that
the IB integrals in question will be dominated by the contributions from
the low-$s$ region. The full VMD-model-based IB contributions, which
provide our estimates for the IB contributions to the $s\le s_{\rm KNT}$,
$3\pi$ integrals, represent $-1.2\%$, $-1.1\%$ and $-1.6\%$
of the corresponding $s\le s_{\rm KNT}$, $3\pi$ totals,
$a_\mu^{{\rm HVP};\, 3\pi}\left( s_{\rm KNT}\right)$,
$I^{3\pi}_{W_{1,5}}\left( s_{\rm KNT}\right)$ and
$I^{3\pi}_{W_{2,5}}\left( s_{\rm KNT}\right)$, respectively.

As noted above, BaBar provides full information on the central values and
errors only for the fit parameters determining the $\rho$, $\omega$ and $\phi$
contributions to the $3\pi$ amplitude, and not for those determining the
$\omega^\prime$ and $\omega^{\prime\prime}$ contributions. The information
that is provided is, however, sufficient to allow us to study what should
be the dominant uncertainties on the estimated IB $3\pi$ contributions
$\left[ a_\mu^{{\rm HVP};\, 3\pi}\left( s_{\rm KNT}\right)
\right]_{\rm IB}^{\rm VMD}$, $\left[ I^{3\pi}_{W_{1,5}}\left( s_{\rm KNT}
\right)\right]_{\rm IB}^{\rm VMD}$ and $\left[ I^{3\pi}_{W_{2,5}}
\left( s_{\rm KNT}\right)\right]_{\rm IB}^{\rm VMD}$, namely
those produced by the uncertainties on the parameters entering the dominant,
fully known, low-$s$ contributions, $\left[ a_\mu^{{\rm HVP};\, 3\pi}
\left( s_\phi \right) \right]_{\rm IB}^{\rho\omega\phi ;\, {\rm VMD}}$,
$\left[ I^{3\pi}_{W_{1,5}}\left( s_\phi \right)
\right]_{\rm IB}^{\rho\omega\phi ;\, {\rm VMD}}$ and
$\left[ I^{3\pi}_{W_{2,5}}\left( s_\phi \right)
\right]_{\rm IB}^{\rho\omega\phi ;\, {\rm VMD}}$.
We find that the dominant uncertainties on these low-$s$ contributions
come from the fit error in the relative phase, $\phi_\rho$, of the $\rho$
and $\omega$ contributions to the amplitude, which produces sizeable
relative uncertainties of $87\%$, $92\%$ and $70\%$ on
$\left[ a_\mu^{{\rm HVP};\, 3\pi} \left( s_\phi \right)
\right]_{\rm IB}^{\rho\omega\phi ;\, {\rm VMD}}$,
$\left[ I^{3\pi}_{W_{1,5}}\left( s_\phi \right)
\right]_{\rm IB}^{\rho\omega\phi ;\, {\rm VMD}}$ and
$\left[ I^{3\pi}_{W_{2,5}}\left( s_\phi \right)
\right]^{\rho\omega\phi ;\, {\rm VMD}}_{\rm IB}$.
There is a smaller $\sim 22\%$ uncertainty in all cases associated
with that on the square root of $B(\rho\rightarrow 3\pi )$, to which the
$\rho$-$\omega$ and $\rho$-$\phi$ interference contributions are
proportional. We expect the uncertainties in the full VMD model estimates
for the total IB contributions resulting from the absence of a
representation of excited $\rho$ resonance effects in the region above
$s=s_\phi$ to be much smaller than the dominant $\phi_\rho$-induced
uncertainties, for the following reason. First, the
contributions to $a_\mu^{{\rm HVP};\, 3\pi}\left( s_{\rm KNT}\right)$,
$I^{3\pi}_{W_{1,5}}\left( s_{\rm KNT}\right)$ and
$I^{3\pi}_{W_{2,5}}\left( s_{\rm KNT}\right)$
from the region between $s_\phi$ and $s_{\rm KNT}$ represent, respectively,
only $6.6\%$, $3.5\%$ and $9.4\%$ of these totals. Even if the omitted
IB $\rho^\prime$-$\omega^\prime$, $\rho^\prime$-$\omega^{\prime\prime}$,
\etc\ interference contributions represented anomalously high $3\%$ fractions
of the total contributions from this region, they would represent only
$0.20\%$, $0.11\%$ and $0.28\%$ of the corresponding $3\pi$ spectral
integral totals, $a_\mu^{{\rm HVP};\, 3\pi}\left( s_{\rm KNT}\right)$,
$I^{3\pi}_{W_{1,5}}\left( s_{\rm KNT}\right)$ and
$I^{3\pi}_{W_{2,5}}\left( s_{\rm KNT}\right)$,
and hence only $17\%$, $10\%$ and $18\%$, respectively, of the
corresponding full VMD model IB estimates
$\left[ a_\mu^{{\rm HVP};\, 3\pi}\left( s_{\rm KNT}\right)
\right]_{\rm IB}^{\rm VMD}$, $\left[ I^{3\pi}_{W_{1,5}}\left(
s_{\rm KNT}\right)\right]_{\rm IB}^{\rm VMD}$ and
$\left[ I^{3\pi}_{W_{2,5}}\left( s_{\rm KNT}\right)\right]_{\rm IB}^{\rm VMD}$,
a level well below that of the dominant low-$s$-region uncertainties. The
$s\le s_\phi$, $\phi_\rho$-induced uncertainties thus strongly dominate
the uncertainties on the estimated IB $3\pi$ double-counting correction,
especially so in the quadrature combination of the above three error
contributions. Since BaBar does not provide the correlations between its
fitted VMD parameters, and a positive $\phi_\rho$-$B(\rho\rightarrow 3\pi)$
correlation would further increase the combined error, we assign, rather
than the quadrature combination, an expanded 100\% uncertainty on our
integrated full VMD-model $3\pi$ IB estimates,
$\left[ a_\mu^{{\rm HVP};\, 3\pi}\left( s_{\rm KNT}\right)
\right]_{\rm IB}^{\rm VMD}$, $\left[ I^{3\pi}_{W_{1,5}}\left(
s_{\rm KNT}\right)\right]_{\rm IB}^{\rm VMD}$ and
$\left[ I^{3\pi}_{W_{2,5}}\left( s_{\rm KNT}\right)\right]_{\rm IB}^{\rm VMD}$.

From the results above, we conclude that it is possible to obtain an
experimentally constrained estimate of the IB $3\pi$ double-counting
correction, albeit with an uncertainty of order $100\%$. This estimate
is of use, despite the sizeable uncertainty, because of the small size
of the central values. Explicitly, the full VMD model IB results,
including this $100\%$ uncertainty, yield
\begin{eqnarray}
\label{ib3piintestimates}
&&\left[ a_\mu^{{\rm HVP};\, 3\pi}\left(s_{\rm KNT}\right)\right]_{\rm IB} =
-0.54(54)\times 10^{-10}\ ,\nonumber\\
&&\left[ I^{3\pi}_{W_{1,5}}\left( s_{\rm KNT}\right)\right]_{\rm IB} =
-0.00041(41)\ ,\nonumber\\
&&\left[ I^{3\pi}_{W_{2,5}}\left( s_{\rm KNT}\right)\right]_{\rm IB} =
-0.00012(12)\ ,
\end{eqnarray}
representing $-1.2\pm 1.2\, \%$, $-1.1\pm 1.1\, \%$ and $-1.6\pm\, 1.6\, \%$,
respectively, of the full $3\pi$ VMD contributions.
As shown in the main text, the uncertainties on these corrections, even
at 100\%, are much smaller than the experimental uncertainties on the
differences of the correspondingly weighted $s=s_\tau$, $2\pi +4\pi$
electroproduction and $\tau$ integrals. The IB $3\pi$ double-counting
correction is thus under good control for the purposes envisioned in
this paper.


\end{document}